\documentclass[aps,twocolumn,superscriptaddress,groupedaddress]{revtex4}
\usepackage{amssymb}
\usepackage{amsmath}
\usepackage{graphicx}
\usepackage{epstopdf}
\usepackage{bm}
\usepackage{soul}
\usepackage{sidecap}
\usepackage[normalem]{ulem}
\usepackage{ mathrsfs }
\usepackage {times}
\usepackage{float}
\usepackage{enumerate}
\usepackage{multirow}
\usepackage{tabularx}
\usepackage{array}
\usepackage{lmodern}
\newcommand\redout{\bgroup\markoverwith{\textcolor{red}{\rule[.5ex]{2pt}{0.4pt}}}\ULon}

\usepackage[colorlinks=true,citecolor=blue]{hyperref}
\hypersetup{colorlinks=true,citecolor=blue,linkcolor=red,urlcolor=blue}

\newcommand{\be}{\begin{equation}}
\newcommand{\ee}{\end{equation}}

\newcommand{\bea}{\begin{eqnarray}}
\newcommand{\eea}{\end{eqnarray}}

\newcommand{\bd}{\begin{displaymath}}
\newcommand{\ed}{\end{displaymath}}
\newcommand{\ba}{\begin{array}}
\newcommand{\ea}{\end{array}}
\newcommand{\bi}{\begin{itemize}}
\newcommand{\ei}{\end{itemize}}
\newcommand{\bc}{\begin{center}}
\newcommand{\ec}{\end{center}}
\newcommand{\bfl}{\begin{flushleft}}
\newcommand{\efl}{\end{flushleft}}
\newcommand{\bfr}{\begin{flushright}}
\newcommand{\efr}{\end{flushright}}
\newcommand{\no}{\nonumber}

\newcommand{\bl}{\begin{aligned}}
\newcommand{\el}{\end{aligned}}

  \def\bd{{\bf d}}

\def\6{\partial}

\def\={\!\!\!&=&\!\!\!}
\def\+{\!\!\!&&\!\!\!+~}
\def\-{\!\!\!&&\!\!\!-~}

\graphicspath{{.}{./Figs/}}

\usepackage{color}
\usepackage[dvipsnames]{xcolor}
\usepackage{xcolor}

\begin{document}
\title{
Disordered Kitaev chain with long-range pairing: Loschmidt echo revivals and  dynamical phase transitions
}

\author{Utkarsh Mishra}\email[]{utkarsh.mishra@apctp.org}
\affiliation{Asia Pacific Center for Theoretical Physics (APCTP), Pohang, Gyeongbuk, 790-784, Korea}
\author{R. Jafari}\email[]{jafari@iasbs.ac.ir}
\affiliation{Department of Physics, Institute for Advanced Studies in Basic Sciences (IASBS), Zanjan 45137-66731, Iran}
\affiliation{Beijing Computational Science Research Center, Beijing 100094, China}
\affiliation{Department of Physics, University of Gothenburg, SE 412 96 Gothenburg, Sweden}
\author{Alireza Akbari}\email[]{alireza@apctp.org}
\affiliation{Asia Pacific Center for Theoretical Physics (APCTP), Pohang, Gyeongbuk, 790-784, Korea}
\affiliation{Department of Physics, POSTECH, Pohang, Gyeongbuk 790-784, Korea}
\affiliation{Max Planck POSTECH/Korea Research Initiative (MPK), Gyeongbuk 376-73, Korea}
\affiliation{Department of Physics, Institute for Advanced Studies in Basic Sciences (IASBS), Zanjan 45137-66731, Iran}

\begin{abstract}
We explore the dynamics of long-range Kitaev chain by varying pairing interaction exponent, $\alpha$.
It is well known that distinctive characteristics on the nonequilibrium dynamics of a closed quantum system are closely related
to the equilibrium phase transitions. Specifically, the return probability of the system to its initial state (Loschmidt echo),
in the finite size system, is expected to exhibit very nice periodicity after a sudden quench to a quantum critical point.
Where the periodicity of the revivals scales inversely with the maximum of the group velocity.
We show that, contrary to expectations, the periodicity of the return probability breaks for a sudden quench to the non-trivial
quantum critical point.
Further, We find that, the periodicity of return probability scales inversely with the group velocity at the gap closing point
for a quench to the trivial critical point of truly long-range pairing case, $\alpha < 1$.
In addition, analyzing the effect of averaging quenched disorder shows that the revivals in the short range pairing cases are more robust against
disorder than that of the long rang pairing case.
We also study the effect of disorder on the non-analyticities of rate function of the return probability which introduced as a witness of the dynamical phase transition.
We exhibit that, the non-analyticities in the rate function of return probability are washed out in the presence of strong disorders.
\end{abstract}
\date{\today}
\maketitle

\section{Introduction}

Recent remarkable advancement of the experimental studies of ultracold atoms, trapped in optical lattices \cite{Bloch2008,Belsley2013}, provide a new
framework for studying the nonequilibrium dynamics of isolated quantum systems, in particular from the viewpoint of quantum quenches \cite{Polkovnikov:2011aa,Mitra:2018aa}.
Quenching a quantum system to/across the critical point raises striking questions, especially, when the time evolution is unitary \cite{Polkovnikov:2011aa,Mitra:2018aa}.
An abrupt change of the state of a closed quantum system leads to a unitary time evolution.
When a sudden quench happens, the evolution is determined by the Loschmidt echo (LE) \cite{Zurek:2005aa}, modulus of overlaps between the eigenstates of
the pre-quenched and post-quenched Hamiltonians expressed by a given change of parameters on which the Hamiltonian depends.
For a sudden quench to a quantum critical point, finite-size case studies reveal that the LE of several models, with short range
interaction, exhibits a periodic revival structure, formed by brief deviation from its mean value \cite{Quan:2006aa,Yuan:2007aa,Rossini:2007aa,Happola:2012aa,Montes:2012aa,Igloi:2011aa,RJHJ2017a,RJHJ2017b}, which can be used as a dynamical witness of quantum criticality \cite{Quan:2006aa,Yuan:2007aa,Haikka:2012aa,Bayat:2015aa,Bayat:2016aa,Bayat:2018aa,Jafari:2015aa,Najafi2017}.

In addition, the nonanalyticities in the rate function of the Loschmidt echo (return probability), when the quench is performed across an equilibrium
quantum critical point, has been lately used to introduce the notion of dynamical phase transitions (DPTs) \cite{Heyl2013,Heyl:2018aa,Vajna:2014aa,Andraschko:2014aa,Kriel:2014aa,Canovi:2014aa}.

Very recently, the studies of the connection between quenching dynamics and quantum phase transition \cite{Zurek:2005aa,Heyl2013,RJHJ2017a}, the topological order \cite{Dimitris:2009aa,Rahmani:2010aa,Halasz:2013aa} and also the dynamics of an edge state  \cite{Bermudez:2009aa,Patel2013,Rajak:2017aa,Sacramento:2016aa}, have attracted the attention of the scientists. Specifically, searching the robustness
and response of the topological edge states to quantum quenches \cite{Rajak:2017aa,Tsomokos:2009aa,Halasz:2013aa,Sacramento:2014aa,Rajak2014}.
The behavior of edge states under a sudden quantum quench has been investigated  in two-dimensional topological insulator \cite{Bernevig1757}, where it was
shown that, in the abrupt transition from the topological insulator to the trivial insulator phase, there is a collapse and revival of the edge states \cite{Patel2013,Sacramento:2014aa}.
Similar results have been obtained for the one-dimensional Kitaev model \cite{Rajak:2017aa,Sacramento:2016aa,Nava2016,Nava2017}.

Current experimental progresses in realization of long-range interacting quantum models with tunable long range interactions \cite{Richerme:2014aa} has renewed the interest in
studying the non-equilibrium dynamics of quantum systems with infinite-range interactions \cite{Zauner:2017aa,Jad:2017aa,Homrighausen:2017aa,Bhattacharya:2018aa,Dutta:2017aa}.
Motivated by the short-range one dimensional Kitaev chain \cite{Rajak:2017aa,Sacramento:2016aa},
a long-range pairing version of an integrable p-wave superconducting chain of fermions has been proposed, where strength of super conducting pairing between two sites
separated by a distance $r$ falling off in a power-law fashion as $r^{\alpha}$  \cite{Viyuela:2016aa,DeGottardi:2013aa, Vodola:2014aa,Hernandez-Santana:2017aa,Van-Regemortel:2016aa,Cai:2017aa,Lepori:2017aa,Alecce:2017aa,Dutta:2017aa, Mahyaeh:2018aa, Ares:2018aa,Li:2018aa,Jad:2017ab}.
Despite numerous attempts to link the significant features of quantum phase transition (QPT) to the quench dynamics (LE), the general principle has not been
established to connect the QPT to dynamics of the systems, specifically in the disordered and long range quantum systems.

In this paper, we study the effects of long range interaction and disorder on the LE of the ground state and edge states of
the long range pairing Kitaev (LRPK) model with open boundary condition. It should be mentioned that the integrability of the LRPK model
breaks in the open boundary condition and also in the presence of disorder.
To the best of our knowledge, such contributions have not been studied in previous works and can shed light on several new effects to the subject.

We show that, in the clean truly LRPK chain, the revival time (periodicity of the revival) in the LE of the finite size system scales
exponentially with the power-law exponent, $\alpha$, and inversely with the group velocity at the gap closing point.
While in the models with short range interaction, as expected, the revivals time is controlled by the maximum group velocity\cite{Happola:2012aa,Montes:2012aa}.
We also show that a surprising result occurs for a quench to the non-trivial critical point where the periodic revivals eliminated and the LE oscillating randomly around its mean value.
Moreover, in the presence of strong local disorder the revivals washed out, and the first revivals in short range pairing cases
are more robust than that of the long range pairing cases.
We further show that, the LE of the localized edge mode in the clean case, for a quench to the critical point, exhibits periodic revivals which increase consequently
with the increase of the power-law exponent, $\alpha$.
These revivals are suppressed in the presence of disorder and disappear for the long-range model while few of them survives in the short-range model for the same strength of disorders.
Finally, studying the dynamical phase transition in the presence of disorder shows that the strong disorder leads to the disappearance of singularity in the rate function of the LE (return probability) \cite{Heyl2013}.

The paper is presented as follows: Sec.~\ref{sec:model} describes the model and its numerically obtained band structure for the open chain. The scheme of global quenching is described in Sec.~\ref{sec:globalquenching} with the techniques to solve the dynamics of the underlying Hamiltonian in the presence and absence of disorder. Dynamics of edge state under sudden quenching is performed in Sec.~\ref{sec:suredge} and the effect of disorder is discussed. Scaling of revival time, obtained from the dynamics of Loschmidt echo in the finite-size system, is reported in Sec.~\ref{sec:revival} including its behavior in presence of disorder and role of power-law pairing exponent. In Sec.~\ref{sec:DQPT}, we present the result on the dynamical phase transition in presence of disorder for long- and the short-range limiting case of the power-law exponent. A discussion on the results is included in conclusion Sec.~\ref{sec:concl}.



\section{The Model}
\label{sec:model}
\begin{figure}[b]
\includegraphics[width=\linewidth]{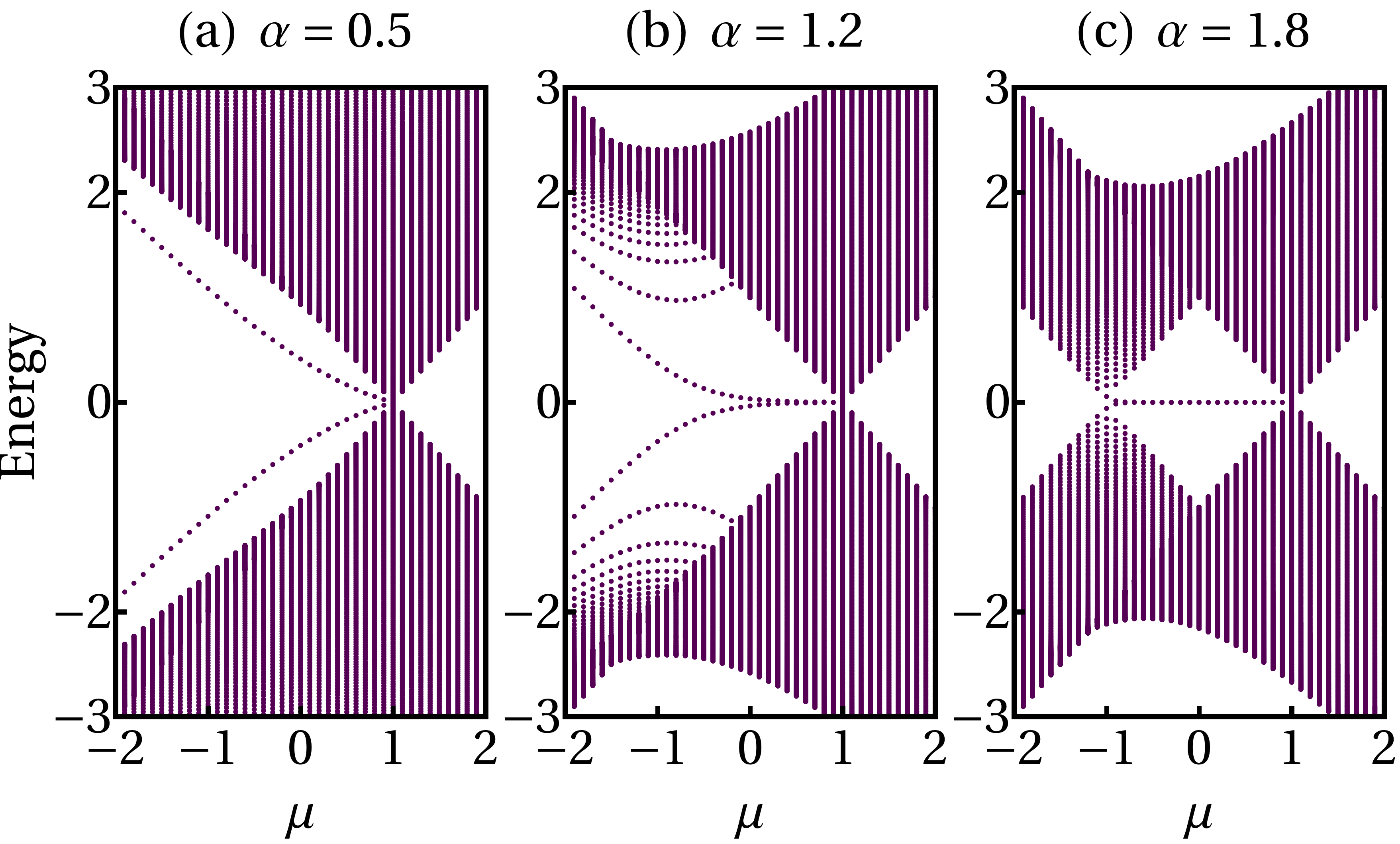}
\\
\vspace{-0.cm}
\hspace{0.2cm}
\includegraphics[width=0.95\linewidth]{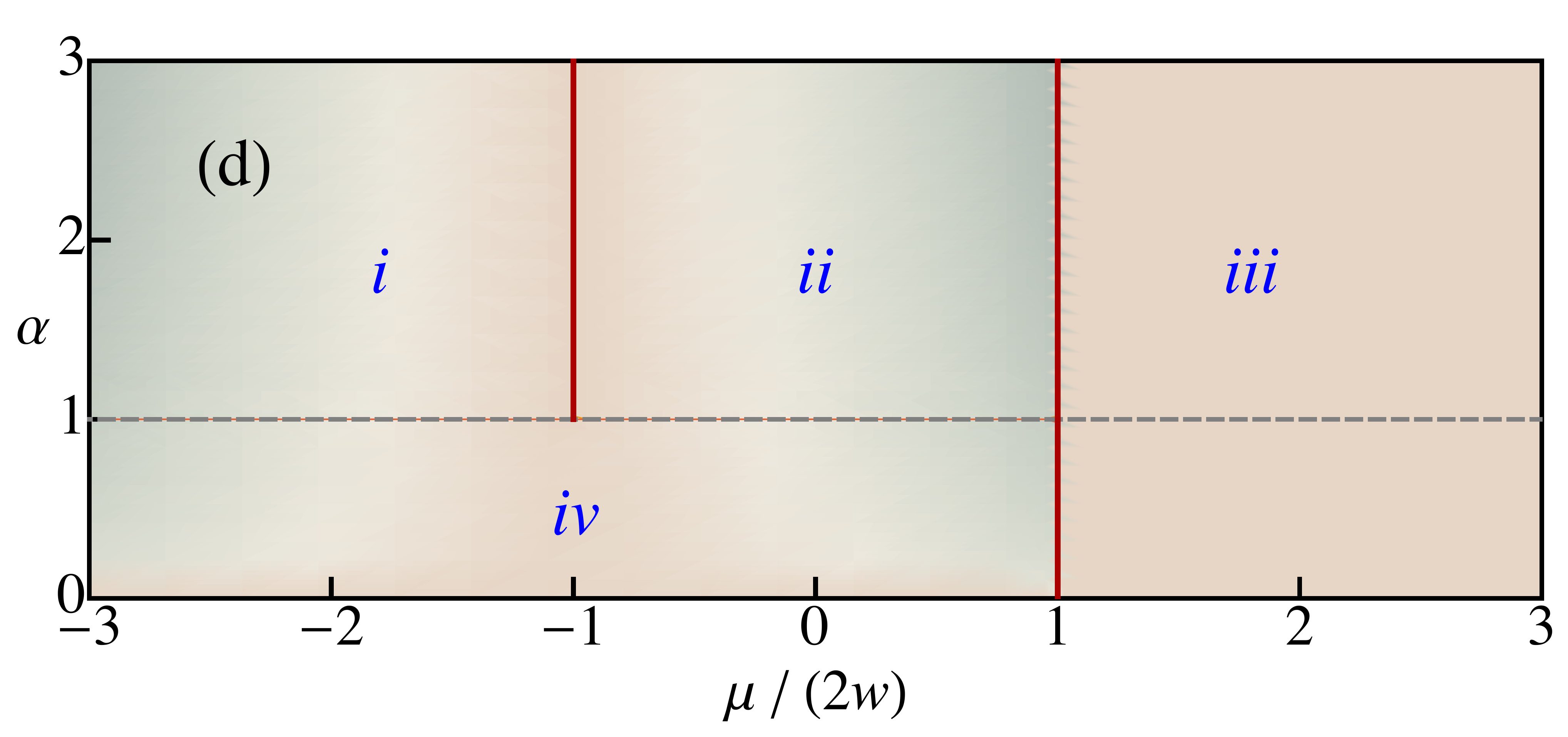}
 \vspace{-0.35cm}
\caption{(Color online)
The energy spectrum of the long-range pairing Kitaev chain with open boundary conditions from an exact diagonalization of the Hamiltonian Eq.~(\ref{eq:LKC}), with respect to varying chemical potential. Here the system size is  $N=500$, $\Delta=2$, and $w=0.5$.
The different panels  are for different value of $\alpha$, namely (a)  $\alpha=0.5$, (b) $\alpha=1.2$, and  (c) $\alpha=1.8$.
(d) Schematic phase diagram of the finite size Kitaev chain with long-range pairing for a different range of the pairing interaction.
The regular solid red vertical lines at $\mu=\pm 1$ denote the gap closing lines.
The dashed gray line in the figure shows the range of true long-range pairing, $0<\alpha \leq 1$.
 }
\label{fig:1}
\end{figure}

\begin{figure*}[t]
\includegraphics[width=\textwidth]{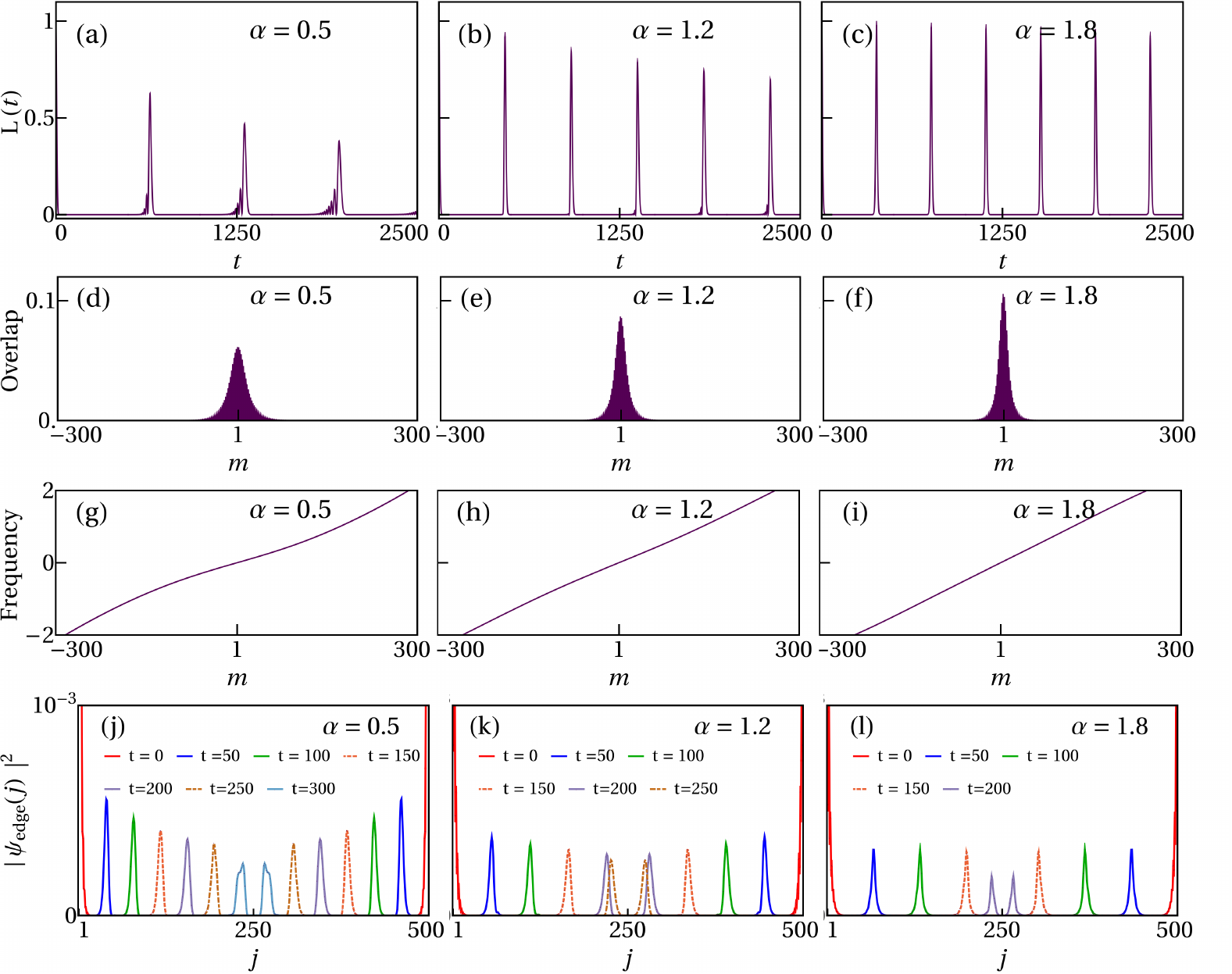}
\vspace{-0.5cm}
\caption{
(Color online)
(a-c) Return probability, $L(t)$, of the long-range Kitaev chain with open boundary condition for critical quenching  from
$\mu_{I}=0.9$ to $\mu_{F}=1$.
(d-f) The overlap of the zero-energy wave function with the wave functions of the final Hamiltonian versus $m$. Here $m$ denotes the indexing of the wave function where wavefunction with negative energy eigenvalues are indexed by negative $m$ and those with positive energy eigenvalues labeled by positive $m$. 
 (g)-(i) The oscillation frequencies, defined as the difference between energy of the initial state with the energies of the final Hamiltonian i.e., $\omega_{m}=E_{n}(\mu_{I})-E_m(\mu_{F})$,  with respect to $m$, for different values of $\alpha$. 
 (h)-(j) Behavior of zero-energy wave-functio, $\psi_{edge}(j)|^{2}$, with respect to lattice sites $j$ at initial time $t=0$ and at various time $t$ after the evolution due to quenching. The depiction is for the critical quenching. The different panel corresponds to different $\alpha$. 
  The system size here is $N=500$, hopping amplitude $w=0.5$, and pairing strength $\Delta=2$.
}
\label{fig:2}
\end{figure*}

We consider long-range pairing Kitaev chain where the pairing interaction is not only present between the nearest-neighbor
sites but at all other distant sites. The Hamiltonian of the model, describing $N-$free fermions in one dimension lattice,  is given by~\cite{Vodola:2014aa}
\begin{eqnarray}
\label{eq:LKC}
\bl
  {\cal H}  =&
   - \sum_{j=1}^{N}
   \Big(
   w(a^{\dagger}_{j}a_{j+1}+H.c.)
   +\mu
   (n_{j}-\frac{1}{2})
   \Big)
   \\
&+ \frac{\Delta}{2} \sum_{j=1}^{N}\sum_{\ell=1}^{N-1}
d^{-\alpha}_{\ell}(a_{j}a_{j+\ell}+H.c.).
\el
\end{eqnarray}
Here, $a^{\dagger}_{j}(a_{j})$ is the fermioninc creation (annihilation) operator on site $j$, $n_{j}=a^{\dagger}_{j}a_{j}$, $w$ is the tunnelling rate, $\mu$ is the chemical potential, and $\Delta$ denotes the strength of the $p$-wave pairing. The summation index $\ell$ varies for each site in the lattice with weightage $d_{\ell}^{\alpha}$. For a closed chain, anti-periodic boundary conditions,
$a_{j+N}=-a_{j}$
makes the effect of long-range paring term intact, with the choice of  $d_{\ell}=\ell$ ($d_{\ell}=N-\ell$) if $\ell\leq N/2$ ($\ell\geq N/2$), respectively~\cite{Vodola:2014aa,Alecce:2017aa}.
For an open chain, $d_{\ell}=\ell$ and we drop terms containing $a_{j>N}$.

A proposal to realize the Hamiltonian of long-range pairing Kitaev chain in the experiment has been put forward recently~\cite{Liu:2018aa}.
For $\alpha\to\infty$, one recovers the standard short-range Kitaev model with pairing range limited to nearest-neighbor.
Therefore, the ground state phase diagram of the Hamiltonian is the same as the Kitaev (Ising spin) model in the limit of $\alpha\to \infty$.
The spectrum of the closed chain depends on the interaction strength $\alpha$. In the thermodynamic limit, the energy gap closes at topological phase transition points $\mu_{c}=\pm1$ for $\alpha>1$.
For $\alpha<1$, in the thermodynamic limit, the bulk degeneracy at the non-trivial critical point $\mu_{c}=-1$ is lifted, while the degeneracy at the trivial critical point $\mu_{c}=1$
remains unaltered. The spectrum of the model has been plotted in Fig.~\ref{fig:1}(a)-(c), for open boundary condition, for different values of $\alpha$, and the Hamiltonian parameters set as
$\Delta=2$, $w=1/2$, and $N=200$.
As seen, the system exhibits an edge state at non-zero energy for $\alpha=0.5$, and $1.2$, separated from bulk in the region $\mu<1$.
For $\alpha=1.8$, the edge state become zero-energy state for $-1<\mu\leq 1$.

A schematic phase diagram of the Hamiltonian, Eq.~(\ref{eq:LKC}), is presented in Fig.~\ref{fig:1}(d).
This model shows a rich phase diagram, with distinct quantum phases $i,ii,iii,$ and $iv$ depending upon the value of $\alpha$ and $\mu$.
The red lines are gap closing lines where the bulk gap closes in the thermodynamic limit.
It is worthwhile to mention that, in finite size system, the bulk gap at $\mu=1$ is very small (closes even for N=40) for any values of $\alpha$,
while along the line $\mu=-1$ the bulk gap is very sizable except for $\alpha\gg 1$ (see the appendix \ref{AppA}).
This completely distress the well-known behaviour of the LE \cite{Happola:2012aa,Montes:2012aa,Jafari:2016aa,Quan:2006aa,Yuan:2007aa} in the finite size system for a quench to the non-trivial
critical points $\mu=-1$.
%
%

\section{Sudden quenching in the long-range Kitaev chain}
\label{sec:globalquenching}
In this section, we elaborate the scheme to study the dynamics of LRPK chain by quenching the chemical potential $\mu$. We consider sudden quenching, where the initial state at time $t=0$ of a pre-quench Hamiltonian, ${\cal H}_{I}(\mu_{I})$, with chemical potential $\mu_{I}$, is evolved under a post quench Hamiltonian,  ${\cal H}_{F}(\mu_{F})$, with chemical potential $\mu_{F}$.  Such global quenching can be realized in current experimental setups~\cite{Jurcevic:2014aa}.
An interesting quantity to look is the return probability \cite{Heyl2013}, the overlap between the initial state and time evolved state under the post-quenched Hamiltonians: 
\begin{eqnarray}
\label{eq2}
{\cal G}(t)=\langle \Psi(\mu_{I})| e^{i  {\cal H}_{I}t}  e^{-i  {\cal H}_{F}t}  |\Psi(\mu_{I})\rangle,
\end{eqnarray}
where $|\Psi(\mu_{I})\rangle$ is the initial state of the system. 
{
The  return probability is known as a Loschmidt echo (LE) defined as modulus of the Loschmidt amplitude
}
\begin{eqnarray}
\label{eq3}
L(t)=|{\cal G}(t)|^{2},
\end{eqnarray}
which is a benchmark of the partial or full reappearance of the original state as a function of time (return probability).
Additionally, the rate function of the return probability, $R(t)$, is defined as
\begin{eqnarray}
\label{eq4}
R(t)=
-\frac{1}{N}
\mbox{Log}
[
 { L}(t)
].
\label{eq:rp}
\end{eqnarray}
If the eigenstate $|\psi_{n}(\mu_{I})\rangle$ of the pre-quenched Hamiltonian $H(\mu_{I})$, with eigenvalues $E_{n}(\mu_{I})$ considered as an the initial state $|\Psi (\mu_{I})\rangle=|\psi_{n} (\mu_{I})\rangle$, then  evolution of the initial state $|\psi_{n}(\mu_{I})\rangle$, under the post-quench Hamiltonian $H(\mu_{F})$, is given by
\bea
\bl
|\psi_{n}(\mu_{I}, t)\rangle=\sum_{m=-N}^{N}e^{i\omega_{m}t}|\psi_{m}(\mu_{F})\rangle\langle\psi_{m}(\mu_{F})|\psi_{n}(\mu_{I})\rangle,
\no
\el
\eea
where $|\psi_{m}(\mu_{F})\rangle$ is the eigenstate of the post-quenched Hamiltonian with corresponding eigenvalue $E_m(\mu_{F})$, and
$\omega_{m}=E_{n}(\mu_{I})-E_m(\mu_{F})$.
Then, the Loschmidt echo amplitude of the initial state $|\psi_{\ell}(\mu_{I})\rangle$ is given by
\begin{eqnarray}
\label{eq5}
{\cal G}(t)=\sum_{m=-N}^{N}e^{i\omega_{m}t}|\langle\psi_{m}(\mu_{F})|\psi_{n}(\mu_{I})\rangle|^{2}.
\end{eqnarray}
Due to the particle-hole symmetry, the number of eigenstates is $2N$. We label the eigenstates with negative energy by $m<0$ and
the eigenstates with positive energy by $m>0$.

The Hamiltonian in Eq.~(\ref{eq:LKC}) can be solved exactly and time dependent quantities defined above can be calculated.  First, Eq.~(\ref{eq:LKC}) can be written in terms of quadratic spinless fermions~\cite{Lieb:1961aa}
\begin{equation}
\label{eq6}
  {\cal H} =\sum_{i,j}a^{\dagger}_{i}{\cal A}_{ij}a_{j}+\frac{1}{2}(a^{\dagger}_{i}{\cal B}_{ij}a^{\dagger}_{j}+h.c.),
\end{equation}
where ${\cal A}$ and ${\cal B}$ are $N\times N$ symmetric and antisymmetric matrices, respectively.
 By defining the operators  $  \Psi^{\dagger}=({\cal C},{\cal C}^{\dagger})=(a_{1},\ldots,a_{N}, a^{\dagger}_{1},\ldots,a^{\dagger}_{N})$   the Hamiltonian can further be written in a simplified form as
 %
 \begin{equation}
  {\cal H} =
\frac{1}{2}
\Psi^{\dagger}
\mathscr{H}
  \Psi;
  \;\;\;
\mathscr{H}=
\begin{bmatrix}
  -{\cal A}  & -{\cal B} \\
   ~ {\cal B}  & ~{\cal A}
\end{bmatrix}.
\end{equation}%
 %
The form of the $N \times N$ matrix  ${\cal A}$ and ${\cal B}$ is obtained from Eq.~(\ref{eq:LKC}).
The Hamiltonian ${\cal H}$ can thus be diagonalized using the unitary operator $U$, given by
\begin{equation}
U=
\begin{bmatrix}
  g  &  h \\
   ~h{^T}  &  g
\end{bmatrix}.
\end{equation}
%
The diagonalized Hamiltonian is given by
\begin{equation}
{\cal H}=\sum_{k}\Lambda_{k}(\eta^{\dagger}_{k}\eta_{k}-\frac{1}{2}),
\end{equation}
%
where $\Lambda$ is the diagonal matrix consists of eigenvalues of
${\cal H}$, and
$(\eta,\eta^{\dagger})^{T}=U({\cal C},{\cal C}^{\dagger})^{T}$, wehere $T$ is the matrix transpose operation.
The elements of the unitary matrix are given by solving the eigenvalue equations:
\begin{equation}
\bl
&
({\cal A}-{\cal B})({\cal A}+{\cal B})\Phi_{k}=\Lambda^{2}_{k}\Phi_{k},
\\ 
&(
{\cal A}+{\cal B})({\cal A}-{\cal B})\chi_{k}=\Lambda^{2}_{k}\chi_{k}.
\el
\ee
%
We solve these equations numerically for finite-size long-range Kitaev chain with open boundary condition. It is then seen that the matrix elements of $g$ and $h$ are obtained as
\begin{equation}
\bl
&
g_{ki}=g_{k}(i)=\frac{1}{2}
\Big(
\Phi_{k}(i)+\chi_{k}(i)
\Big) ,
 \\&
h_{ki}= h_{k}(i)=\frac{1}{2}
 \Big(\Phi_{k}(i)-\chi_{k}(i)
 \Big).
\el
\ee
%
%
For the Hamiltonian in Eq.(~\ref{eq:LKC}), ${\cal G}(t)$ can be written as~\cite{Rossini:2007aa}
%
\begin{eqnarray}
{\cal G}(t)=\mbox{Det}[{\cal I}-M+M e^{-i   {\cal H}_{F}t}].
\label{eq:mmatrix}
\end{eqnarray}
%
Here ${\cal I}$ is the identity matrix and we have
%
\begin{equation}
M=
\begin{bmatrix}
   \langle {\cal C}^{\dagger}   {\cal C}\rangle  &  \langle {\cal C}^{\dagger}{\cal C}^{\dagger}\rangle  \\
     \langle {\cal C}\;{\cal C}\rangle  &  \langle {\cal C}\;{\cal C}^{\dagger}\rangle
\end{bmatrix}.
\end{equation}%
%
The $2N\times 2N$ matrix $M$, defined in this way corresponds to the correlation matrix of the initial state
$|\psi_{\ell}(\mu_{I})\rangle$.
More precisely,  $ \langle {\cal C}^{\dagger}   {\cal C}\rangle =h^{T}h$ and
$\langle {\cal C}^{\dagger}{\cal C}^{\dagger}\rangle=h^{T}g$. It is to be noted that the conservation of particle number implies
$ \langle {\cal C}^{\dagger}   {\cal C}\rangle+\langle {\cal C}{\cal C}^{\dagger}\rangle= {\cal I}$.

%
%
\begin{figure}
\vspace{-0.05cm}
\hspace{-0.085cm}
\includegraphics[width=0.495\textwidth]{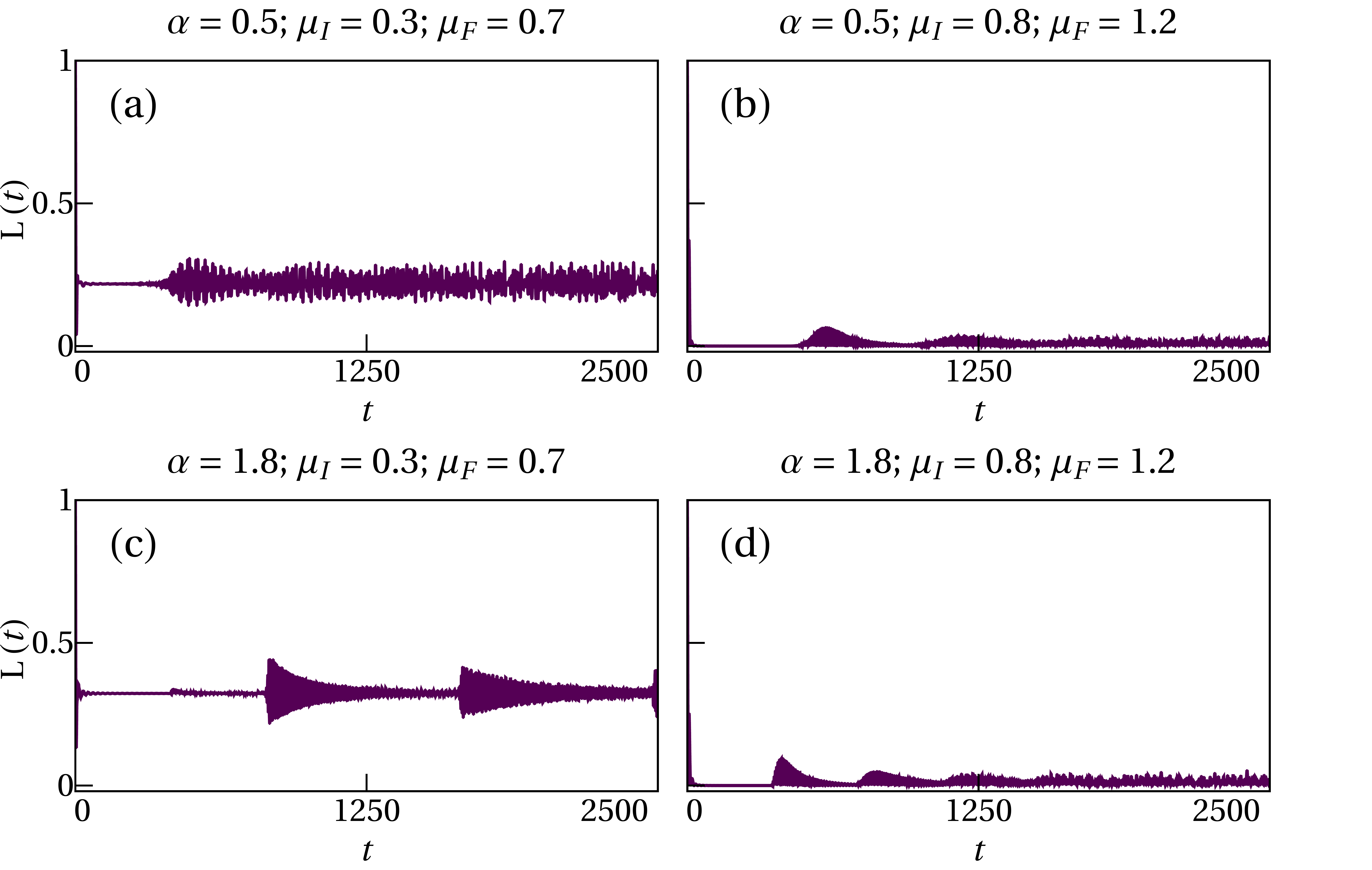}
\\
\includegraphics[width=0.475\textwidth]{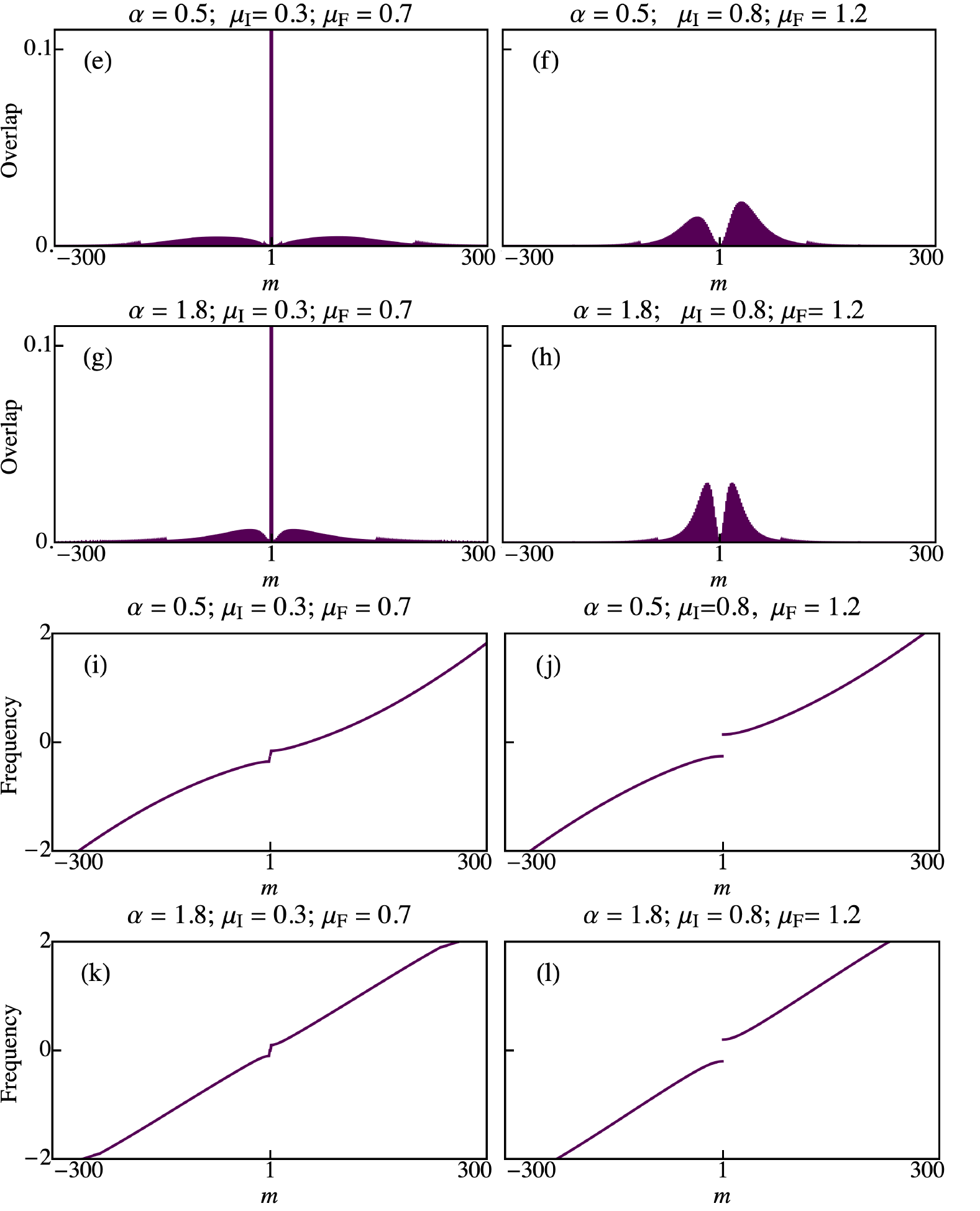}
\vspace{-0.35cm}
\caption{(Color online)
(a-d) Return probability, $L(t)$,  of the zero-energy mode in the long range Kitaev chain with open boundary condition.
(e-h) describe the behavior of the overlap function with respect to the energies of the final Hamiltonian, where $m$ is the index of the energy levels.
Panels (a,~c),  (e,~g) and (i,~k) are representing the quenching within the same phase, whereas (b,~d), (f,~h), and (j,~l) are for the process across the critical phases.
(i-l) The oscillation frequencies (the variation in the difference of energies of the final Hamiltonian and the energy of the initial state) as a function of $m$.
In all the panels, the system size is $N=500$ and other Hamiltonian parameters are same as in Fig.~\ref{fig:2}.
}
\vspace{-0.2cm}
\label{fig:3}
\end{figure}
%


\section{Quantum Revivals in the Loschmidt echo of the Edge State}
\label{sec:suredge}
Localized edge states are a distinctive feature of many-body quantum systems with topological phases         \cite{Hasan:2010aa}. In the Kitaev chains, such localized edge modes are identified as Majorana modes  \cite{Kitaev:2001aa}. The robustness of Majorana modes under certain perturbation of system parameters is an interesting topic to investigate. We here analyze the dynamics of edge (zero energy) states of the long-range pairing Kitaev chain with varying pairing interaction exponent. Two types of quenching are considered: (i) critical quench, i.e., quenching to the critical point and (ii) non-critical quench, i.e., quenching away from the critical point.
The role of disorder on the stability of edge states under quench dynamics to the critical point is also presented.
We should remind the fact that the system possesses distinct topological features depends on the values of $\alpha$ (See Fig.~\ref{fig:1})~\cite{Bhattacharya:2018aa}. 
%
\subsection{Non-disordered Chain}
The return probabilities, $L(t)$, of zero energy eigenstates of the pre-quench Hamiltonian, localized at the edges of the chain, are shown in  Figs.~\ref{fig:2}(a)-(c) for the quenching the system from $\mu_{I}=0.9$ to the critical point $\mu_{F}=1$. 
The return probabilities $L(t)$ exhibit periodic revivals with respect to time with decreasing amplitudes of the subsequent revivals. 
They clearly show that the revival time of the edge states depends on the power-law exponent of LRP interaction, 
and the amplitude of the revivals enhances with an increase of $\alpha$.
Furthermore, as expected, Eq.~(\ref{eq5}), the return probability of the system to the initial state depends upon the overlaps of initial state with the eigenstates of
the post-quenched Hamiltonian, $A_{m}=|\langle \psi_{{m}}(\mu_{F})|\psi_{n}(\mu_{I})\rangle|^{2}$. It also depends on the difference between the initial state's energy of the pre-quenched Hamiltonian and eigenstates' energy of the post-quenched Hamiltonian $\omega_{m}$, which we termed oscillation frequencies henceforth.

From a mathematical point of view, due to the oscillatory behavior of ${\cal G}(t)$ all terms contribute destructively
in the Loschmidt amplitude to cancel each other which results randomly oscillating LE.
The only terms that can survive in the Loschmidt amplitude to contribute
significantly in the return probability $L(t)$, are those with large overlaps function (oscillation amplitude), $A_{m}$.
In addition, since we are interested in the periodic revivals in the LE, according to the discrete Fourier transform,
the ${\cal G}(t)$ reveals periodic behaviour if the oscillation frequencies change linearly with respect to $m$, over the range [$-m_{cut},m_{cut}$] where the overlap functions are non-zero. In other words, $\omega_{m}\approx lm\omega_{1}$ where $l\in (0,1]$.
The overlap functions, $A_{m}$, are plotted in Fig.~\ref{fig:2}(d)-(f) with respect to $m$ for different values of $\alpha$.
As seen, the overlaps are peaked around the edge states $m=\pm1$. In fact, the significant contribution to the overlaps achieves by
the energy states of the bulk around the zero energy (energy of the initial state).
Since overlaps function can be interpreted as measuring the probabilities of particle excitations, the energy states around
the zero energy point (bulk gap closing point) are indeed expected to result much larger overlaps. So, the energy states nearby
the zero energy point contribute sizably in Eq.~(\ref{eq5}).
As $\alpha$ increases the amplitude of overlaps curve enhances, which results in large revival amplitude [See Fig.~\ref{fig:2}(b)~and~(c)].
It means that reappearance probability of short range system in its initial state (edge state) is more than that of the long range case.
The oscillation frequencies, $\omega_{m}$, are plotted in Fig.~\ref{fig:2}(g)-(i) with respect to $m$ for different values of $\alpha$.
As seen, $\omega_{m}$ changes linearly with $m$ around the zero energy states (nonzero overlaps) and the linearity gets perfect as
$\alpha$ increases.
To better understand the physics behind the presence/absence of the revivals in the LE, we probe the spectrum of the model.
The bulk energy gap is very tiny at $\mu=1$ for all values of $\alpha$, which results notable overlaps around the zero bulk energy gap. 
While the results exhibit small overlaps for $\mu_{F}=-1$ arises because of sizable bulk energy gap at $\mu=-1$ (see Fig.~\ref{fig:APP2} in appendix~\ref{AppB}).
This result indicates that the appearance of revivals at the finite size system, when the initial state is edge state, is controlled by the energy states around the bulk gap closing point.

In Fig.~\ref{fig:2}(h)-(j), the time evolution of the Majorana states is depicted for the critical quench.
As is clear, when the system quenched to the critical point $\mu_{F}=1$, separating two topological phases, the edge states oscillate
between two ends of the chain and each edge cross each other in a solitonic-like behavior.
The time period of this oscillation is proportional to the size of the chain.
In addition, the amplitude of edge states decoherences with time,
due to the interference of bulk states of higher energy with the edge states as they oscillate back and forth between the two ends.
%
Since the information propagates through the system via the wave packets of quasiparticles, the bulk gap closing point can be interpreted as the
reference point where all quasiparticles are synchronized and the revival time can be interpreted as the time instances at which all quasiparticles
are synchronized with the bulk energy gap closing mode \cite{RJHJ2017a,Jafari:2018aa,Jafari:2018ab}.
\\
%

\begin{figure}[t]
\includegraphics[width=1.05\linewidth]{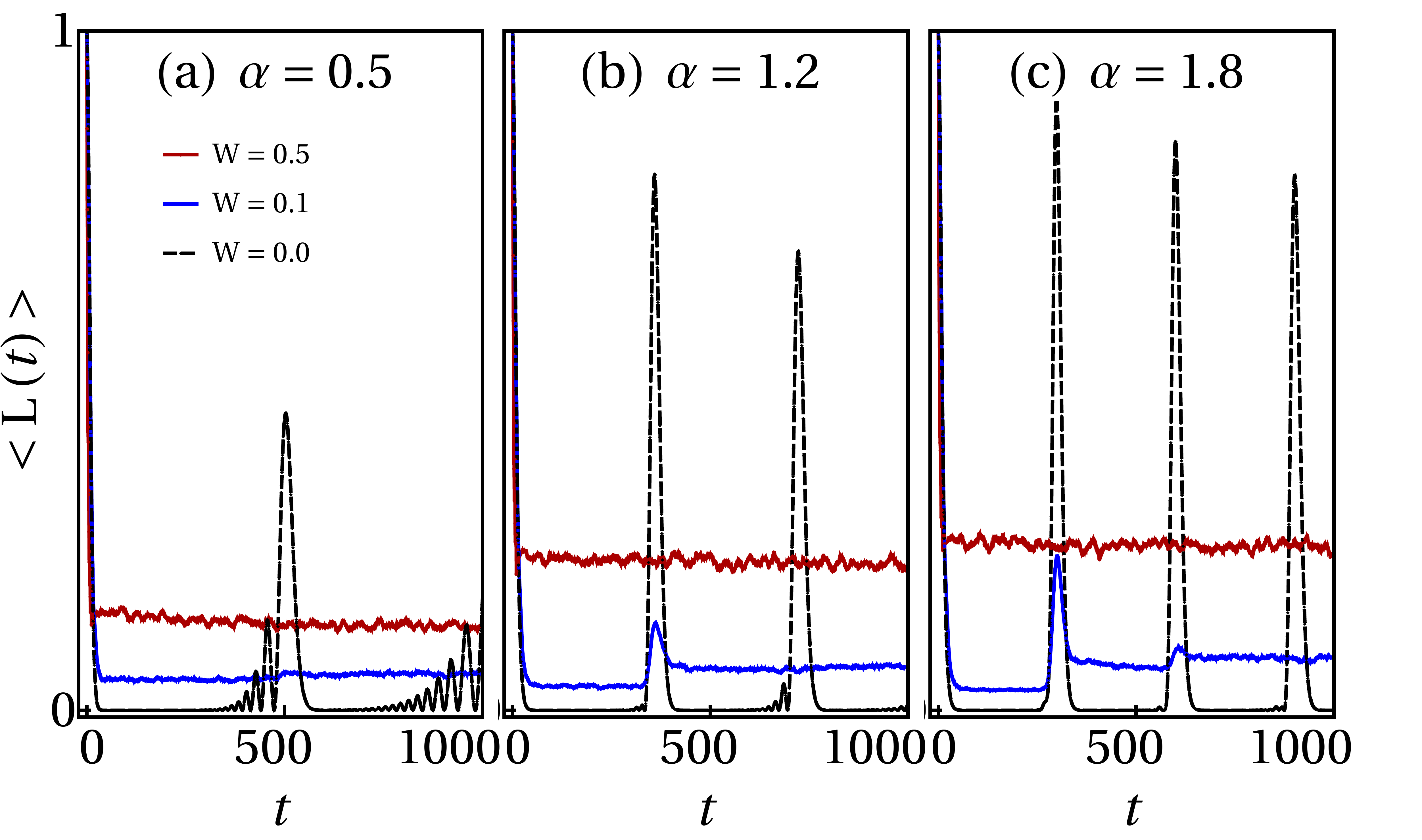}
\vspace{-0.5cm}
\caption{(Color online)
Effect of disorder on the return probability of zero-energy modes. The initial wave function is  zero-energy edge mode for a non-disordered system. The disorder is introduced in the post quench Hamiltonian. The different curves in the panel are for two disorder strengths $W=0.1,0.5$ respectively. The return probability of the clean case ($W=0$) is also shown by dashed lines. 
The ordinate denotes the disorder average  of the return probability, $L(t)$, over $50$ disorder samplings of $\mu_{j}=\mu+V_{j}$
with $V_{j}\in [-W,W]$.
The parameters set as
 $N=200$, $\mu_{I}=0.9$, and $\mu_{F}=1$. The other parameters of the Hamiltonian is same as in Fig. \ref{fig:2}. 
}
\label{fig:4}
\end{figure}

\begin{SCfigure*}
\includegraphics[width=0.7\textwidth]{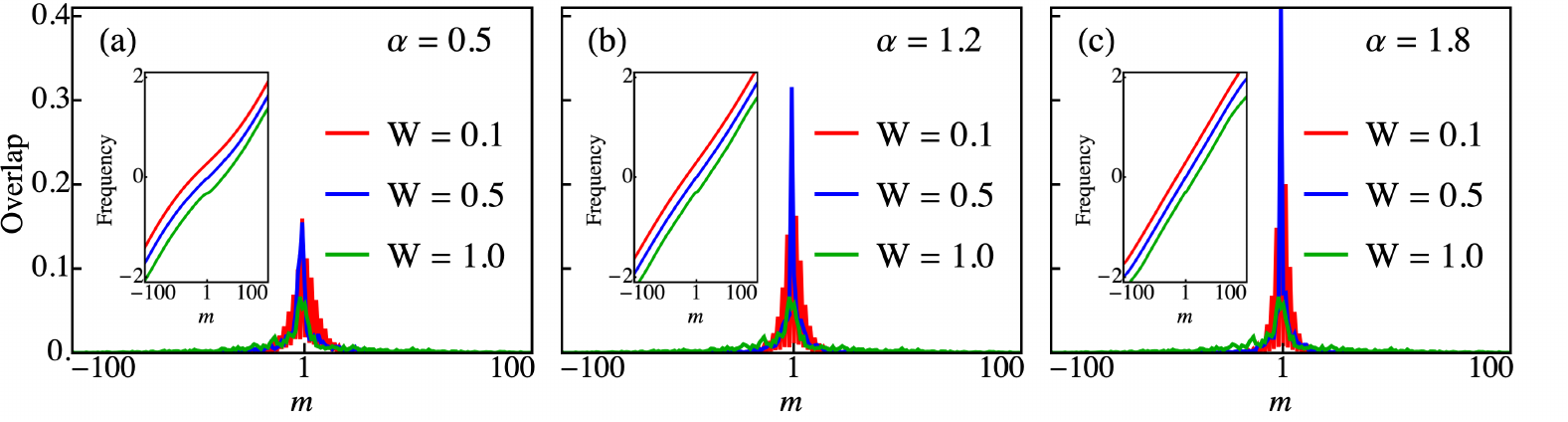}
\vspace{-0.1cm}
\caption{(Color online) (a-d) show the overlap of the zero energy wave-function with respect to $m$, for the different strength of disorder which is represented by $W$.  The inset shows the behavior of the energy difference (for better visualization, the vertical axis were shifted by $\pm 0.3$ for red and green lines).    }
\label{fig:5}
\end{SCfigure*}

In Fig.~\ref{fig:3}(a)-(d), we have plotted the return probabilities of the edge state for quenching to the non-critical
point for various cases of initial and final parameters.
In particular, we consider two cases of non-critical quenching: (i)  quench in the same phase, and (ii) quench across the critical point.
The results show that for both types of non-critical quench the $L(t)$ decays rapidly with time and remains small with noisy and small amplitude oscillation.
%
This behavior can be understood from the overlap functions and oscillation frequencies. 
From Figs.~\ref{fig:3}(e)-(h) one can clearly see that the overlaps are very small away from the zero energy
and the peak at zero energy has a single point for a quench within the same phase. Further, the oscillation frequencies are not linear with respect to $m$ in Figs. \ref{fig:3}(i)-(j) while it is linear in Figs. \ref{fig:3}(k)-(l).
As discussed, the LE does not show revival for the non-critical quench due to the small overlaps and nonlinear oscillation frequencies.
It is remarkable to mention that, the quasi-revivals in the LE in Figs.~\ref{fig:3}(c)-(d)
arises from the linear behavior of the oscillation frequencies around the zero energy.
Moreover, the single point peak in the overlaps of the quench within the same phase results in the larger mean value, than that of quench across the critical point.



\subsection{Disordered Chain}
\label{subsec:disLE}

The quench disorder is a crucial factor responsible for affecting the behavior of physical quantities in many-body systems as compared to the clean case~\cite{Mishra:2016ab}.
Therefore, the robustness of given physical quantities, in our case LE,  can be established by incorporating the effect of disorder in the Hamiltonian~\cite{Happola:2012aa}.
We now consider random chemical potential at all the sites of the form $\mu_{j}=\mu+V_{j}$ in Eq.~(\ref{eq:LKC}). Here $V_{j}$  are random numbers taken from a uniform distribution with widths $W$, i.e., $V_{j}\in [-W,W]$. Thus $W$ here denotes the strength of the disorder.
In Fig.~\ref{fig:4}, the effect of disorder on the return probability is analyzed for different $\alpha$, for a quench to the critical point
$\mu_{F}=1$. For each time $t$, a finite number of disorder-sampling is drawn randomly and quench disorder averaging is performed over these finite samples to calculate the average time dependent probability $\langle L(t)\rangle$. It can be seen that the revival amplitudes in $L(t)$, suppressed due to disorders.
For small disorder, $W=0.1$, there appears only one revival in the time spam of $t\in[0,1000]$, in the disordered case as compared to the two revivals in the clean case, Fig.~\ref{fig:4}(b). Similarly, in Fig.~\ref{fig:2}(c) there are three revivals while the presence of disorder leads to suppression of one revival for $\alpha=1.8$, as shown in Fig.~\ref{fig:4}(c). For $\alpha=0.5$, there appear no revival due to disorder, while in case of strong disorder, $W=0.5$, in Fig.~\ref{fig:4}(b-c), the revivals in $\langle L(t)\rangle$, washed away. Note that the time of revival does not get changed in the presence of disorder in the return probability. Interestingly, the mean value of return probability is greater in the case $W=0.5$, as compared to the small disorder strength $W=0.1$. It is to be noticed that, the amplitude of revival for $\alpha=1.8$, is higher than that of $\alpha=1.2$.
As a consequence, in the presence of disorder, the edge state revivals in the LE of short range interaction case are more robust than that of long range interaction model.

In Fig. \ref{fig:5} the overlaps function and oscillation frequencies have been plotted versus $m$. As seen, the height of overlaps function is increased as disorder increases which result in higher return probability's mean value for strong disordered cases. On the other hand, the oscillation frequencies deviate from the linear behavior as the disorder gets stronger. The nonlinear behavior of the oscillation frequencies depresses the periodic behavior of the LE in the presence of the disorders.


\begin{flushright}
\begin{figure*}[t!]
\includegraphics[width=0.321\textwidth]{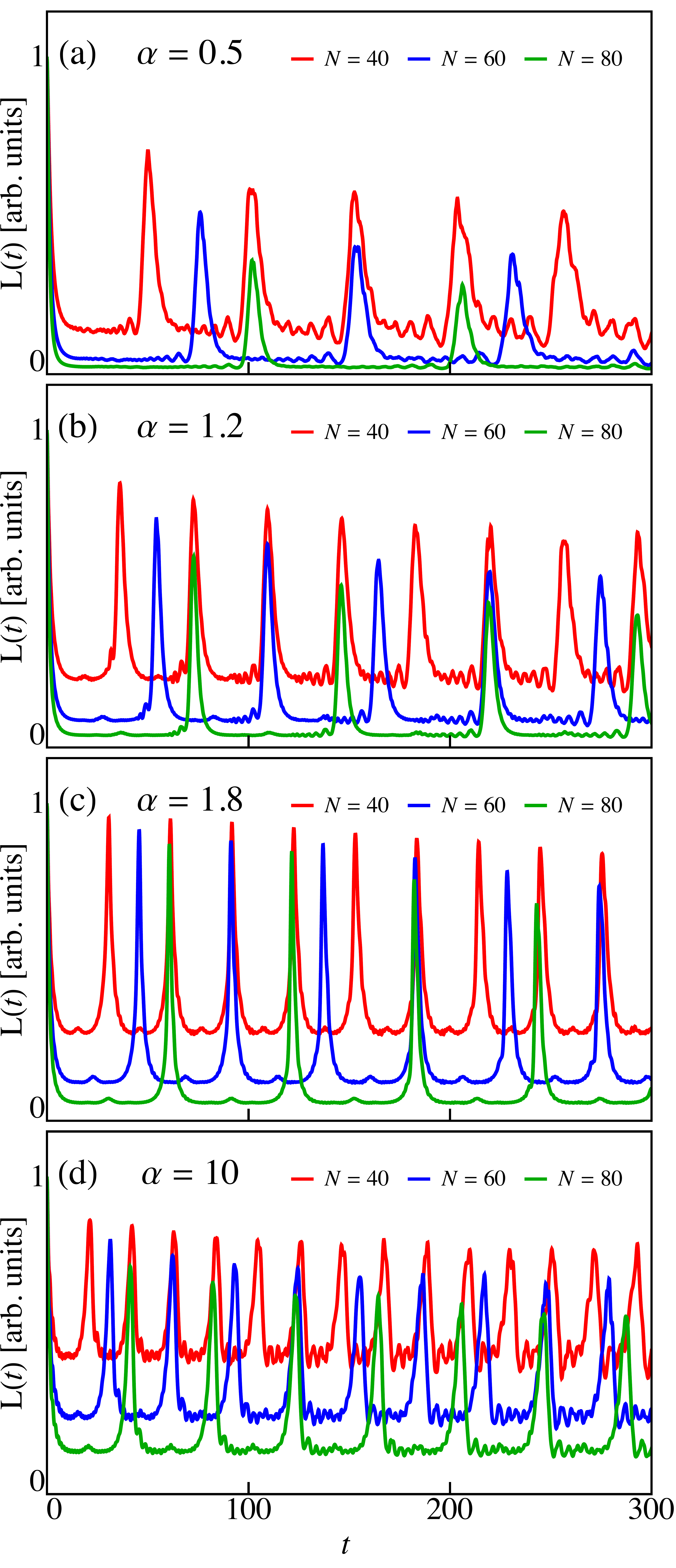}\hfill
\includegraphics[width=0.33\textwidth]{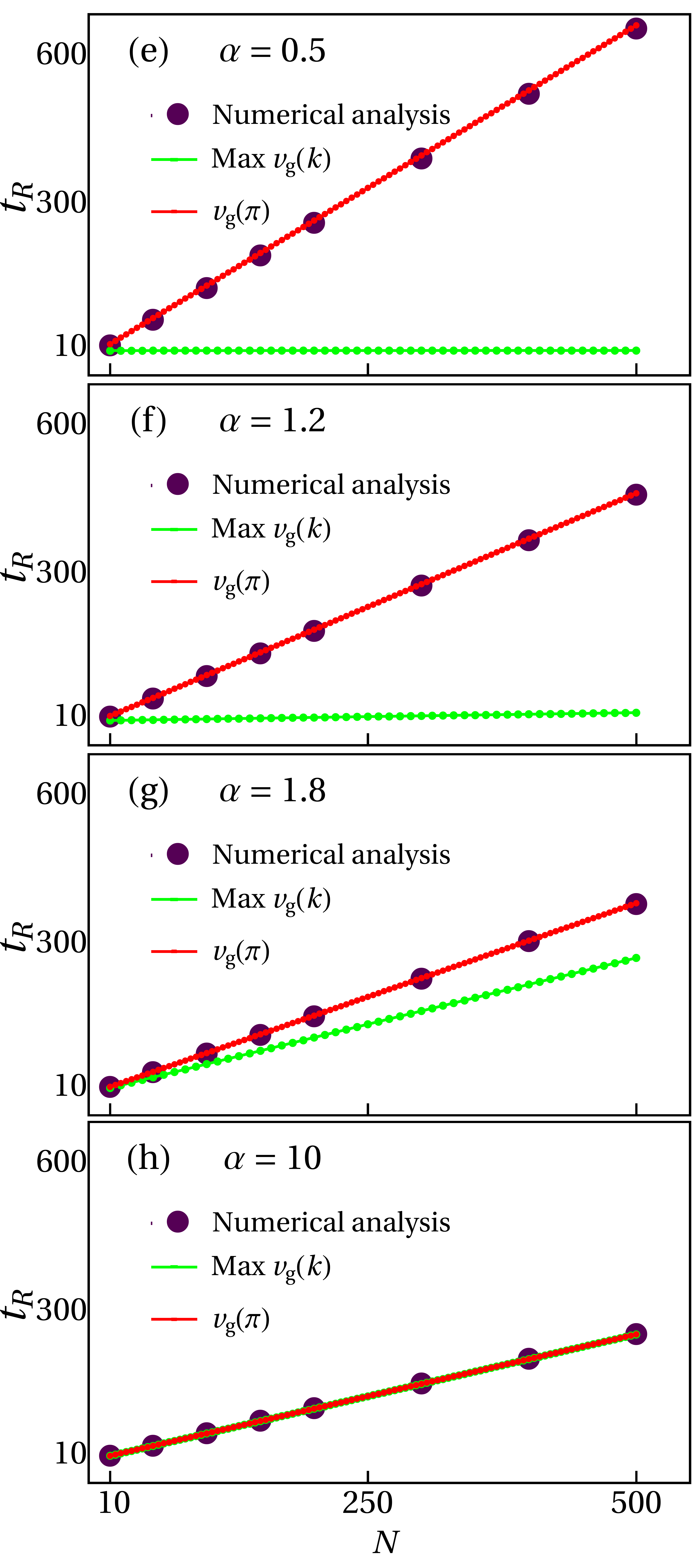}\hfill
\includegraphics[width=0.337\textwidth]{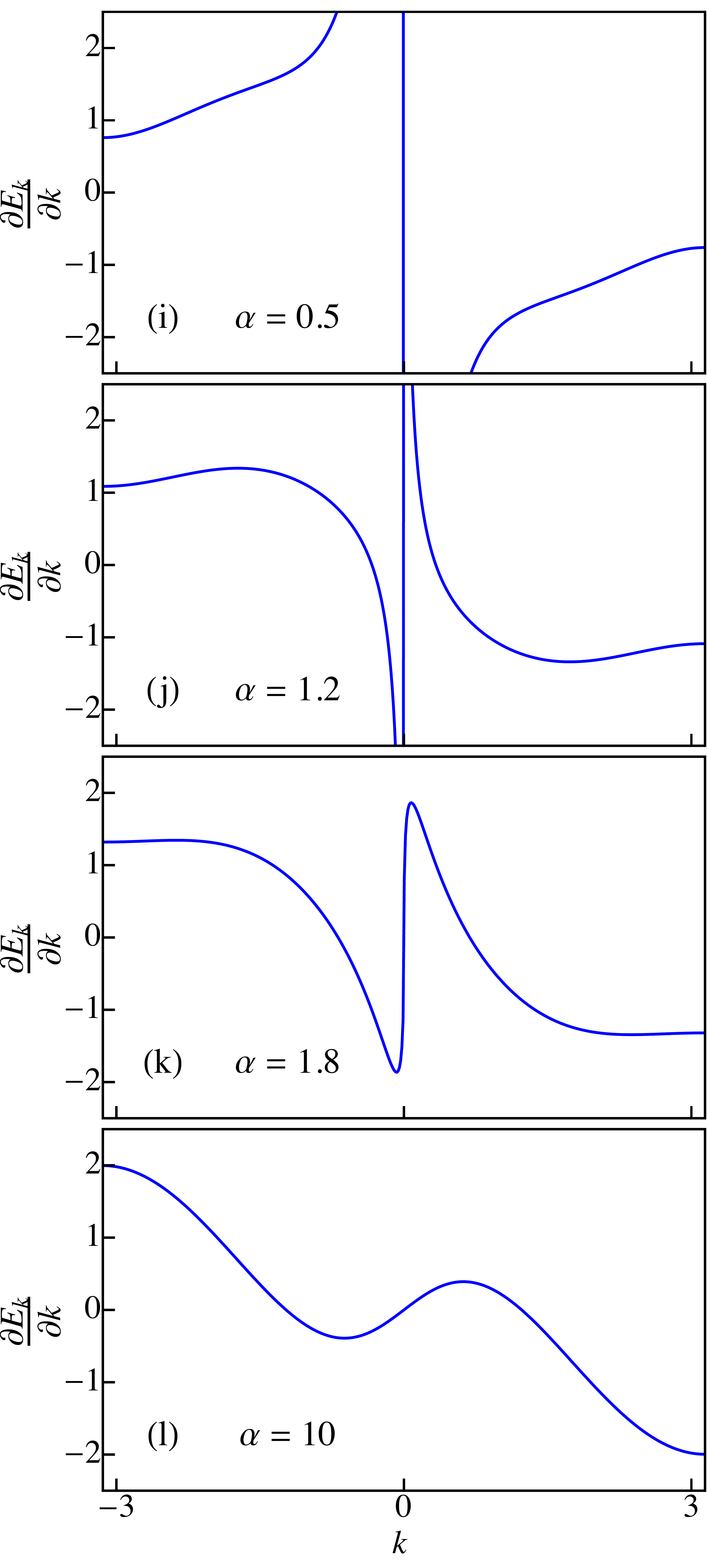}
\vspace{-0.2cm}
\caption{(Color online)
 (a-d): The evolution of the Loschmidt echo, $L(t)$, as a function of time, $t$, for different system size due to the global quenching in the long-range Kitaev chain with open boundary condition (OBC). The red, blue, and green lines are for $N=40,60$, and $N=80$, respectively.
 (e-h) The plot of the first revival time $t_{R}$ vs system size for the different strength of pairing ranges (different $\alpha$ values).
 The purple filled circles denote data of the first revival time obtained from (a-d), and the fitted data are shown by solid lines, in which
 the red and green curves are obtained by fitting them with equation $t_R=N/v_g$ using the group velocities ($v_g$) at the momentum corresponds to the gap closing mode $k=\pi$ and where it takes the maximum value, respectively.
  Notice that, as shown in (h),  for the short-range pairing ($\alpha\gg 2$), all the fitting curves merge together matching well to the numerical data (filled circle).
 (i-l) The variation in the derivative of energy spectrum $E_{k}$ with respect to $k$ for various choices of $\alpha$.
 Here, we set the Hamiltonian parameters as $\Delta=2$, $w=0.5$, $\mu_{I}=1.2$ and $\mu_{F}=1$.
 }
\label{fig:6}
\end{figure*}
\end{flushright}
\section{Quantum Revivals in the Loschmidt echo of the Ground State}
\label{sec:revival}
In this section, we investigate dynamics of the ground state in both clean and disordered LRPK chain after a sudden quench. We consider a finite size system with open boundary conditions and quench the system, prepared in its ground state for given $\mu_{I}$, to the critical point of the post-quenched Hamiltonian, i.e., $\mu_{F}=1$.
Our aim, in this section, is to explore the dynamics of Loschmidt echo in the different topological regions by considering the ground state critical quenching.

\subsection{Non-disordered Chain}
We show the properties of Loschmidt echo for four values of $\alpha$ belonging to the different ranges in  Fig.~\ref{fig:6}. The Loschmidt echo decays rapidly with time for a short period of time and reaches a stationary value. This signature becomes prominent with increasing the system size, while as expected, for small size system there are fluctuations in the LE around the stationary value. This feature can be seen from Figs.~\ref{fig:6}(a)-(d). 
After a lapse of certain time $t$, the Loschmidt echo revives and reaches to a maximum value. The time $t$ when the LE reaches to the maximum is called the first revival time $t_{R}$ as it is the first instance during the evolution when the LE reaches to a value close to the initial value at $t=0$. The rapid decay of LE in the short time scale and periodic revivals are also shown in Appendix \ref{AppB} for few cases of $\alpha$ but for higher system size, $N=500$. In the present section, however, the results are for system size $N=40,60,80$, as it makes analysis accessible through plots.   
From Figs.~\ref{fig:6}(a)-(d), it is clear that the revivals of the LE depend on the system size. In Fig.~\ref{fig:6}(e)-(h), we plot the first revival time $t_{R}$ with respect to the system size $N$. The first revival time of Loschmidt echo in many-body quantum system can be approximately given by $t_{R}\approx N/v^{max}_{g}$ (see~Refs. [\onlinecite{Happola:2012aa,Montes:2012aa}] for the details), where $v^{max}_{g}$ is the maximum group velocity.
In our case, the group velocity $v_{g}$ is a function of the power-law decaying pairing interaction given by exponent $\alpha$. A numerical calculation of the derivative of energy spectrum $E_{k}$ with respect to $k$ provides a plot of the group velocity as a function $k$, as shown in Figs.~\ref{fig:6}(i)-(l).
Once we obtain the group velocity, $v^{\alpha}_{g}$ at the critical point, the revival time can be approximated using the above formula by replacing the group velocity for given $\alpha$. The green regular line in Figs.~\ref{fig:6}(e)-(h) is the approximation of the revival time using maximum group velocity, denoted as Max $v_{g}(k)$ in the plots. It is clear Figs.~\ref{fig:6}(e)-(h) that the data of the revival time obtained by exact calculation do not correspond to the predicted revival time with Max $v_{g}(k)$.
Next, we consider the group velocity at the gap closing mode $k=\pi$, denoted as $v_{g}(\pi)$ in the plots. The revival time with this velocity is well approximated as shown in blue regular line in Figs.~\ref{fig:6}(e)-(h). Surprisingly, both the curves of analytical revival time merged into one for $\alpha\gg 1$.
%
\begin{figure}[t]
\vspace{-0.45cm}
\includegraphics[width=1.0\linewidth]{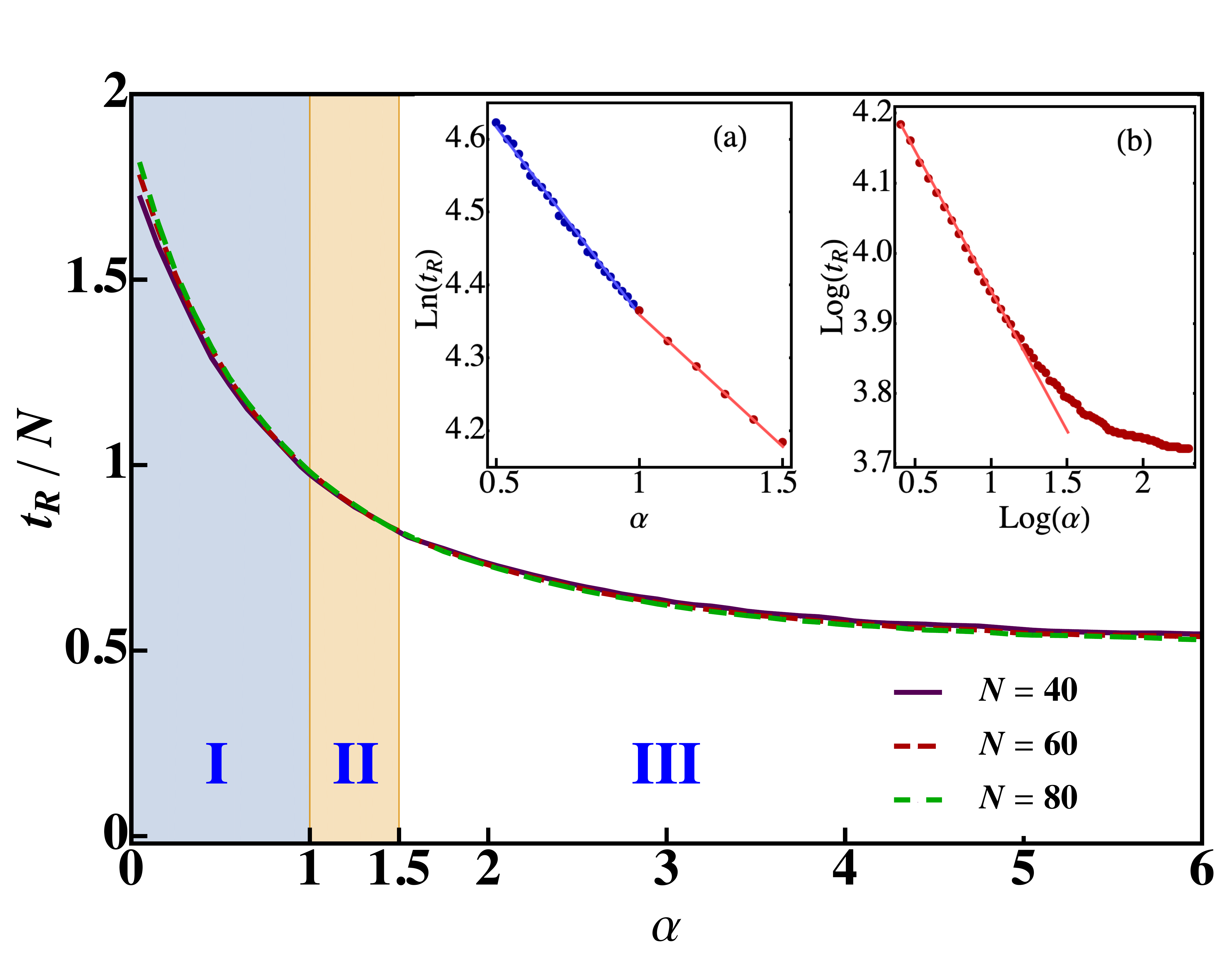}
\vspace{-0.75cm}
\caption{(Color online)
The revival time $t_{R}/N$ with respect to interaction strength $\alpha$ for three considerable  system sizes  $N=40,60,$ and  $N=80$.
Here  $\Delta=2$,  $w=0.5$, $\mu_{I}=1.2$, and $\mu_{F}=1$, and the open boundary condition is assumed in the calculation.
The different limiting regions  are labeld by I for  $ \alpha \in (0,1)$,  II  for $\alpha \in (1,1.5]$,  and III for  $\alpha \in[1.5,\infty]$, respectively [see Fig.~\ref{fig:1}(d)].
Note: for $\alpha \to \infty$, $t_{R}$ approaches to the value of the short-range limit. 
The  inset (a) represents that the first revival time $t_{R}$ decays exponentially in the region I \& II, 
and the inset (b) shows that the decay in the region III is polynomial.
}
\label{fig:7}
\end{figure}
%
\begin{figure}[t]
\includegraphics[width=1.01\linewidth]{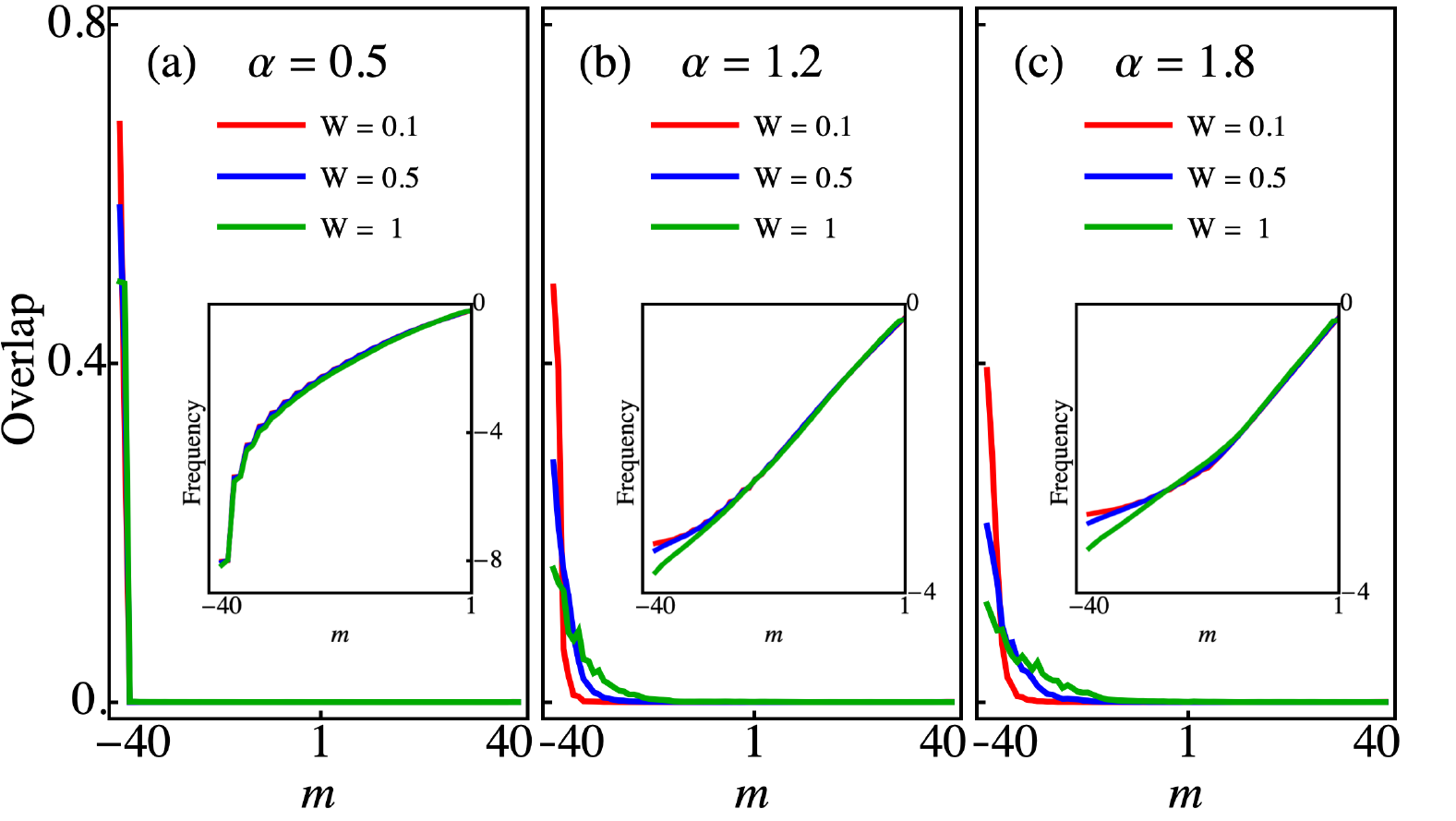}
\vspace{-0.75cm}
\caption{(Color online) Overlap of the initial ground state with the wave-function of the final Hamiltonian. Each such wave function is indexed by $m$. The different panels are for different values of $\alpha$, $(a)\alpha=0.5$, $(b)\alpha=1.2$, and $(c)\alpha=1.8$. The insets show the behavior of frequency $E_{m}(\mu_f)-E_{0}(\mu_I)$, where $E_{0}$ is the ground state of the initial Hamiltonian.  Here system size is $N=40$, and the other parameters are same as in Fig.~\ref{fig:6}.  }
\label{fig:8}
\end{figure}
By fixing the system size, and varying the interaction strength $\alpha$, we can find the dependence of $t_{R}$ on the interaction strength $\alpha$. In Fig.~\ref{fig:7}, we plot the normalized revival time, $t_{R}/N$, with respect to $\alpha$. The plots show collapse of data for different system sizes $N=40$ (regular purple line), $N=60$ (dashed red line), and $N=80$ (dashed double-dashed green line). For better analyzing the behavior of revival time with respect to $\alpha$, we divide the effect  of alpha into three different ranges
(I)    $\alpha\in (0,1)$,
(II)   $\alpha \in (1,1.5]$, and
(III)  $\alpha \in[1.5,15]$.
The  first revival time $t_{R}$ decays exponentially in the range $\alpha\in (0,1)$ and $\alpha\in(1,1.5]$ (left inset Fig. ~\ref{fig:7}).
The rate of decay is faster in the region I as compared to region II. The decay of $t_{R}$ is polynomial in the region III as shown in right inset, Fig.~\ref{fig:7}. For $\alpha\to \infty$, the $t_{R}$ approaches to the value of the short-range Kitaev chain.
In particular the decay of $t_R$ follows the scaling functions: $t_{R}\sim e^{-0.6 \alpha}$ in region I, $t_{R}\sim e^{-0.3\alpha}$ in region II, and  $t_{R}\sim \alpha^{-0.3}$ in region III.

We have also performed quenching to the non-trivial critical point $\mu_{c}=-1$ in the finite size LRPK wire. Contrary to expectations, the results show that periodic revivals are absent for quenching to the critical point $\mu_{c}=-1$ for $\alpha>1$ except very large $\alpha$.
To understand the origin of different behaviors of the LE at both the critical points, $\mu=\pm 1$, let us recall Eq. (\ref{eq5}).
%

%
As discussed, the energy states of the post-quenched Hamiltonian which have energy quite close to the ground state's energy of the pre-quenched Hamiltonian is indeed expected to result in considerable overlaps.
Namely, the states with energy very close to the energy of the initial state can satisfy both conditions, large overlaps and linear variation in oscillation frequencies simultaneously.
Consequently, the LE is expected to exhibit periodic revivals at time instances over which all states with energy close the initial state's energy contribute constructively in Eq. (\ref{eq5}).
%

%
As seen in Figs.~\ref{fig:8}(a)-(c), the overlaps are delta function type at the energy point where is close to the energy of the initial state for all values of $\alpha$. While, the results exhibit very small overlaps for a quench to the non-trivial critical point $\mu_{F}=-1$, except for very large values of $\alpha$.
This result indicates that the appearance of revivals in the finite size system is controlled by the energy states of post-quenched Hamiltonian which have a strong resemblance to the initial state.
%


\subsection{Disordered Chain}

\begin{figure}[t]
\includegraphics[width=1.04\linewidth]{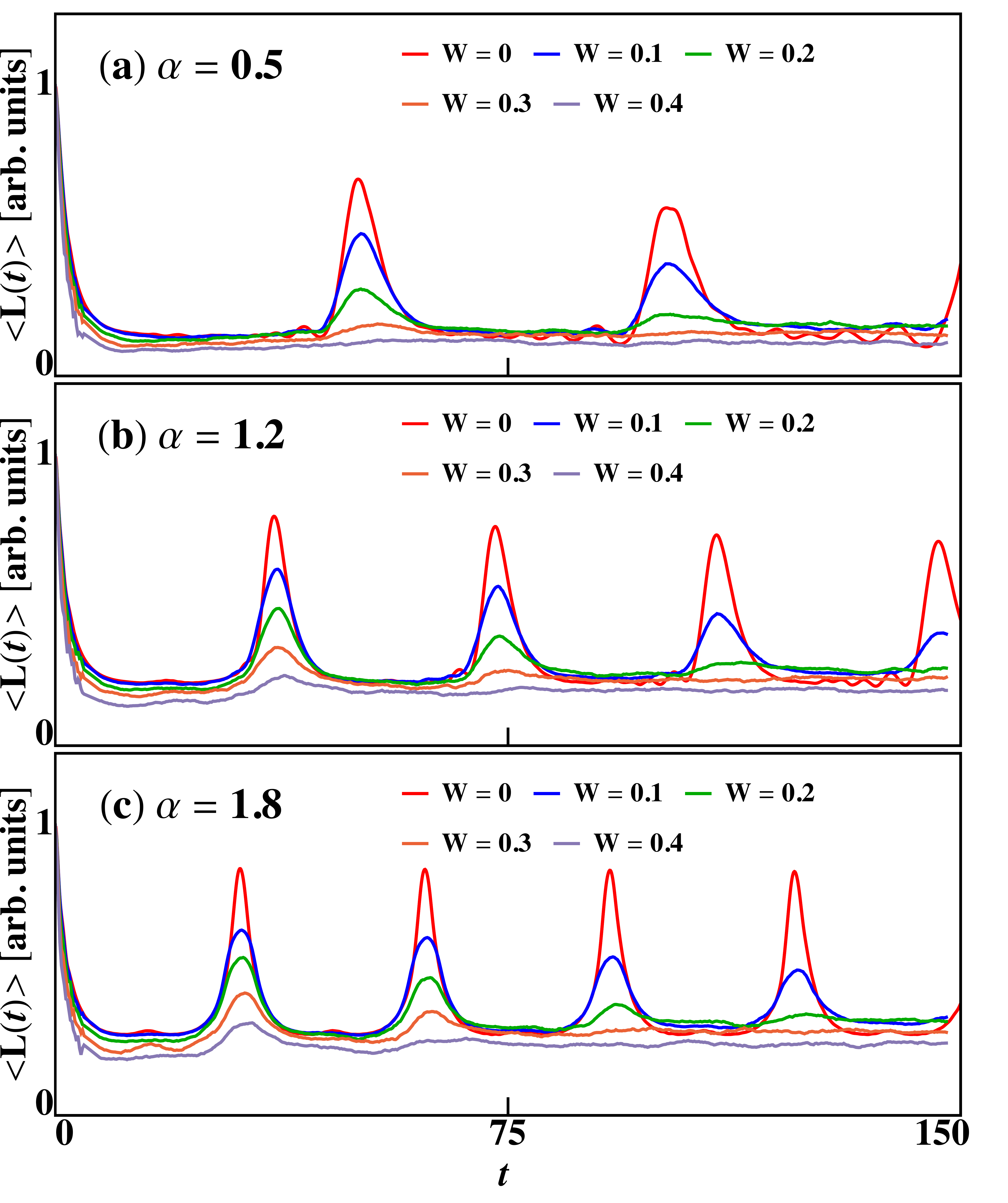}
\vspace{-0.75cm}
\caption{(Color online)
Variation of the Loschmidt echo, $L(t)$, with respect to time for different disorder strengths, $W$, for  (a)  $\alpha=0.5$, (b)  $\alpha=1.2$,   and (c)   $\alpha=1.8$.
 The disorder is present in the chemical potential, $\mu_{j}=\mu+V_{j}$, with $V_{j}\in [-W,W]$.
The system size is $N=40$,
and we consider open boundary condition, and the quenching parameters are same as in Fig.~\ref{fig:6}. The ordinates denotes the average Loschmidt echo over different disordered realizations.  }
\label{fig:9}
\end{figure}

The behavior of the LE is shown in Fig.~\ref{fig:9} for LRPK chain with open boundary condition and system size $N=40$ for different $\alpha$. We consider disorder chemical potential $\mu_{j}$, at each site $j$, of the Hamiltonian Eq.~(\ref{eq:LKC}) as described in subsection \ref{subsec:disLE}. The Loschmidt echo here is calculated over $50$ different disorder realizations for each time. Finally, quench average LE is presented in the plots where the averaging is denoted as $\langle L(t)\rangle$.
In all the figures, it is clear that the LE decreases by increasing the strength of disorder which means the probability of reappearance of
the system in the initial state decreases by increasing the strengths of disorder. Although, the first revival time $t_{R}$ in the presence of
the disorder is the same as that of the non-disordered chain, the disorder affects the periodicity of revivals for disorder strength $W\geq 0.3$.
The presence of disorder in the system for small $\alpha$ shows that the first revival in the LE suppressed for disorder strength $W\geq 0.3$, Fig.~\ref{fig:9}(a). While for the same disorder strength the revival survives for $\alpha=1.2$ and $1.8$, Figs.~\ref{fig:9}(b)-(c).
Finally, they completely suppressed in the presence of strong disorder.
This means the revivals due to the disorder are more robust in short range case than the long range model.

To comprehend how the disorder affects the revivals in the LE we have plotted the overlaps function in presence of disorder for different values of $\alpha$ (Fig.~\ref{fig:8}).
As seen, the disorder disrupts the energy levels and overlaps amplitude around the ground state energy decreases sizably by increasing the strength of disorder. Suppression of revival in the presence of disorder can be attributed to the reduction of overlaps between initial state and eigenstates of post-quenched Hamiltonian.
Moreover, since the overlaps of a few states are nonzero, a few oscillation frequencies contribute to the LE.
Table \ref{table1} displays the oscillation frequencies corresponds to the non-zero overlaps in Fig.~\ref{fig:8}(a)-(c) for different values of disorders.
\begin{table}[h!]
\centering
\caption{The oscillation frequencies corresponds to very large overlaps in Fig.~\ref{fig:8}(a) for different values of disorders.}
\begin{tabular}{ |c || c|c|c|c| }
\hline
$W$ & $~~0~~$ & $0.1$  & $0.5$  & $1$
\\
\hline
 $\omega_{1}$ & $-0.088 $ &   $-0.120$ &  $-0.128$  & $-0.150$
\\
\hline
 $\omega_{2}$ & $-0.172 $  &  $-0.180$ &  $-0.195$  & $-0.198$
  \\
\hline
\end{tabular}
\label{table1}
\end{table}
%

As it is clear, for non-disorder case $\omega_{2}\approx2\omega_{1}$ while in the presence of disorder oscillation frequencies deviates from linearity $\omega_{m}\neq(m+1)\omega_{0}$.
In summary, the probability of resemblance of lower energy states of the post-quenched Hamiltonian to the initial state decreases in the presence of disorder.
%


\section{Dynamical phase transition in disordered long-range pairing Kitaev chain}
\label{sec:DQPT}
Dynamical phase transition is referred to the situation when the return probability, Eq.~(\ref{eq:rp}), 
exhibit singularity at some time $t$. This analogy stem from the equilibrium partition function in complex temperature plane, first pointed out by Yang and Lee~\cite{Yang:1952aa,Lee:1952aa}. The equilibrium transition point corresponds to the crossing of Yang-Lee zeros to the real temperature axis of the complex partition function~\cite{Fisher:1965aa,Yang:1952aa,Lee:1952aa}. Similarly, the complex time, $z=t+i \tau$, provides an analogous picture where the zeros of  ${\cal G}(z)$ cross the real time axis at time $t=t_{c}$ referred as the critical time~\cite{Heyl2013}. It is claimed that when this crossing happens, the return probability $R(t)$ shows a singularity at that time $t$.

The presence of local disorder on the dynamical phase transition has been considered in a very few studies~\cite{Obuchi:2012aa,Takahashi:2013aa, Yang:2017aa, Yin:2018aa}. In this paper, we consider the situation, when the chemical potential of the pre-quench Hamiltonian is fixed at $\mu_{I}$ and the chemical potential of the final Hamiltonian is chosen randomly from a uniform distribution around fixed $\mu_{F}$. The choice can be managed numerically by considering a set of $\mu_{F}$, as $\{\mu_{F}\}=\langle \mu_{F}\rangle+[-W,W]$, where $W$ is the width of the disorder and $\langle \mu_{F}\rangle$ is the mean value of the final quenching which is same as the $\mu_{F}$ of the clean system, i.e., with $W=0$. This setting allows us to investigate the effect of disorder on the dynamical phase transition of the clean system with final chemical potential $\mu_{F}$.  It is akin to the effect of disorder on the equilibrium phase transitions in quantum systems~\cite{Sachdev:2011aa}.  The impact of disorder is not a very well understood concept in the nonequilibrium phase transition.  In the next few paragraphs, we will investigate the impact of the disorder on the nonequilibrium phase transition of long-range pairing Kitaev wire for two different values of the power-law decaying pairing exponent, $\alpha$.

\begin{figure}[t]
\vspace{-0.6cm}
\hspace{-0.5cm}
\includegraphics[width=1.04\linewidth]{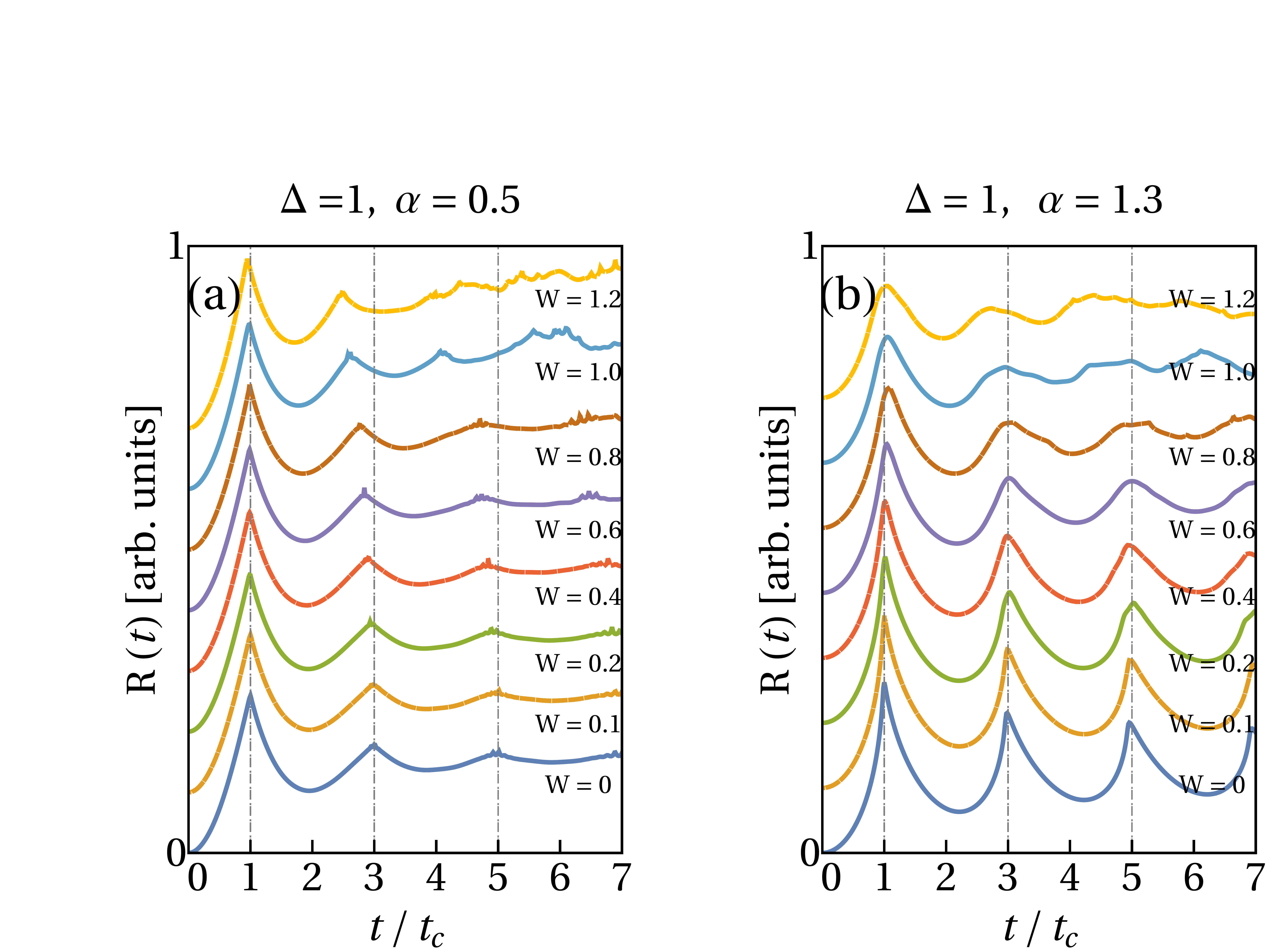}\\
\vspace{-0.46cm}
\hspace{-0.6cm}
\includegraphics[width=1.04\linewidth]{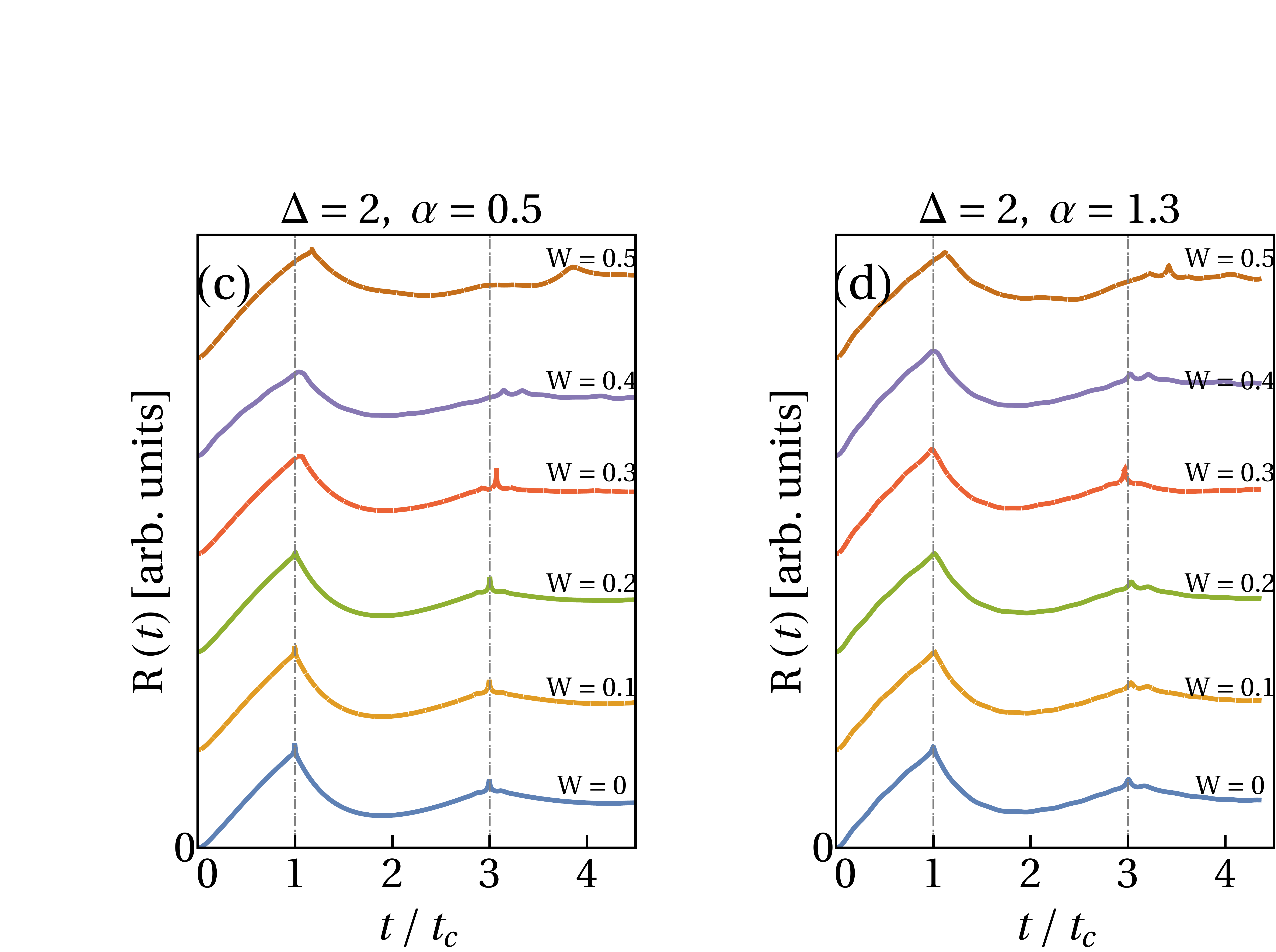}
\vspace{-0.25cm}
\caption{(Color online)
The rate of return probability, $R(t)$,
of   long-range Kitaev chain for various disorder strengths
 and two choices of alpha: (a and c) $\alpha=0.5$ and
(b and d)  $\alpha=1.3$. The time 
is rescaled by $t_{c}$ which is the first instances   singularity appear in $R(t)$.
The parameters set as $N=240$, $\mu_{I}=0.7$, and $\mu_{F}=1.3$.
The panel (a and b) are for $\Delta=1.0$, and panel
(c and d) are for $\Delta=2.0$.
The plot with the individual disorder is shifted by a constant value along the $y$-direction for better visibility.
}
\label{fig:10}
\end{figure}

In Fig.~\ref{fig:10}(a), we show the return probability for various disorder strengths, $W$, for quenching to gap closing point $\alpha=0.5$ and $\Delta=1.0$. In clean long-range Kitaev chain, the return probability displaying singular behavior at times $t=(2n+1)t_{c}$, where $n=0,1,2,\ldots$~\cite{Heyl:2018aa}.
  For disorder system, the chemical potential for each local sites of a total system size of $N=240$, are drawn from a uniform distribution as $\{\mu_{F}\}=1.3+[-W,W]$ with initial parameter $\mu_{I}=0.7$. Thus the quenching $\mu_{I} \to \mu_{F}$ is such that it crosses the gap closing point. For low disorder, the critical time remains the same as the clean one. The significant impact of the disorder on the dynamical phase transitions starts visible at higher disorder strengths $W>0.4$, where the time at which  $R(t)$  diverges shifts to lower time. The curves in Fig.~\ref{fig:10}(a) are shifted by a constant value along the ordinate for clear visibility of the pattern at individual disorder. Without this shift, all the curves get overlapped at the first critical time. We further investigate the effect of disorder by considering the disorder strengths $W\gg1$ and observed that the singularity at $t_{c}$ starts disappearing with high disorder strengths. It is also be noted that here we consider single-instant of disorder realization with maximum width of $W$. Taking disorder averaging does not change the observed behavior significantly and the qualitative behavior of the plots remains unaffected with disordered averaging. In Fig.~\ref{fig:10}(b), we consider the value of $\alpha=1.3$ and investigate the impact of disorder on the dynamical phase transition of the clean system. The clean LRPK chain show dynamical phase transition at $t=(2n+1)t_{c}$, where $n=0,1,2,\ldots$, marked by the singular behavior of return probability $R(t)$.  Focusing on the first critical time occurs at $n=0$, it can be seen from Fig.~\ref{fig:10}(b) that the singularity in $R(t)$ goes away and the transition becomes smooth as the disorder strength $W$ is increased. 
 Smoothening of the singularity also develops at later critical times where the dynamical phase transition is present with zero or small disorder strength as can be seen from Fig.~\ref{fig:10}(b). To check the consistency of the effect of disorder, we also consider $\Delta=2.0$ in Fig.~\ref{fig:10}(c-d). We again observe that for increasing the disorder strengths, the singularity in $R(t)$ washed away. It is to be mentioned here that the above claim is also valid for larger system size. We presented the analysis for $N=500$ in Appendix \ref{AppC}. 


\section{Conclusion}
\label{sec:concl}
We report the dynamics of return probability by considering two choices of the initial state of evolution: (i) localized edge state (ii) ground state of the clean and disorder long-range pairing Kitaev chain. The effect of disorder in the dynamics of localized edge state shows that the oscillations in the survival probabilities are present for large $\alpha$ while they are absent in the small $\alpha$. We also present the survival dynamics of localized edge modes for different quenching. It is found that quenching within the same phase conceal the survival of localized edge state while the same get enhanced for quenching across the gap closing point. A possible explanation of the observed phenomenon is also provided. 
The dynamics of quantum revivals in the return probability, from the ground state, is periodic for a quench to the gap closing point and the periodic structure become more transparent for this quenching. We found that the prediction of the first revival time is best approximated using the ansatz $t_{R}\approx N/v_{g}$, where $v_{g}$ is taken at the gap closing mode. A naive approximation of $v_{g}$ with maximum group velocity, on the other hand, fails to predict the correct revival time for small $\alpha$ (long-range). Thus, we provide the relevant time scale present in the long-range Kitaev chain which captures the correct time of the first revival of LE.
 Moreover, for quenching to the critical point where gap does not close in a finite chain, there exist no revivals in the LE.
  We further investigate the effect of disorder on the return probability and find that the time of the first revival is stable even in the presence of small disorder strength in the system. The periodic structure of the return probability, on the other hand, are affected by the presence of disorder. 
Finally, we consider the dynamical phase transition in the model and the effect of disorder for two different cases of pairing effect. The singularity in rate function of return probability survives in small disorder strengths while the curve becomes smother for the large disorder.
We believe that our results shed further insight into the dynamics of clean and disordered long-range pairing Kitaev wire. The results presented here can also be simulated in experiments with the help of the present state of the art in experimental setups in cold-atoms and ion-traps.

\begin{figure*}
\vspace{0.45cm}
\includegraphics[width=0.95\textwidth]{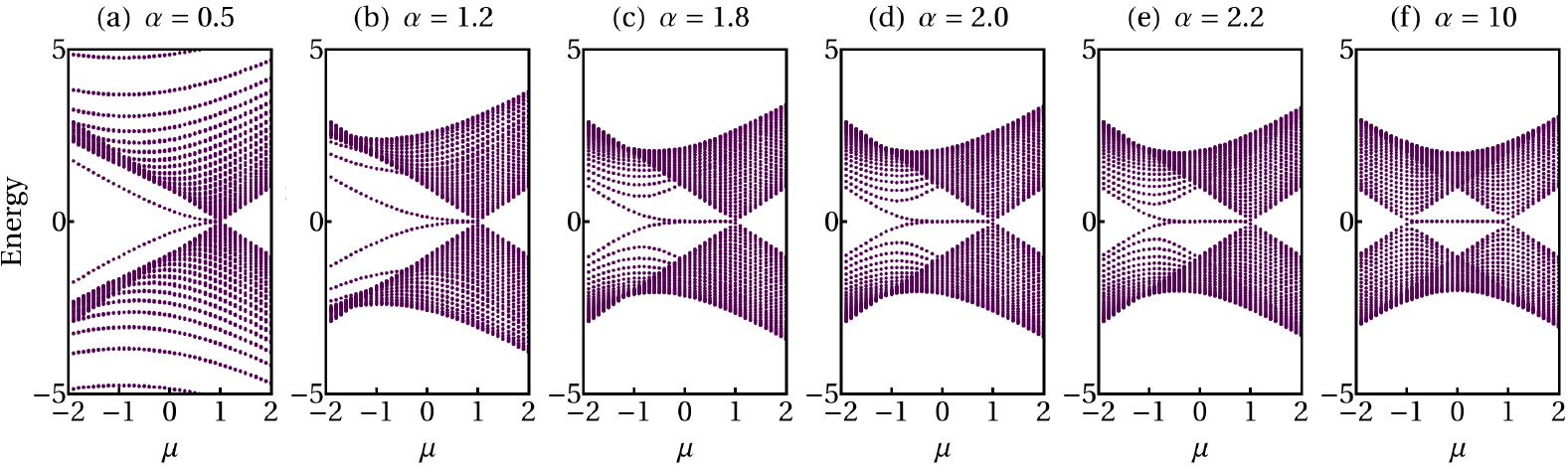}
\caption{(Color online) Energy spectrum of the long-range pairing Kitaev chain as a function of onsite energy $\mu$ for different pairing exponents $\alpha$: $(a)=0.5$, $(b)=1.2$, $(c)=1.8$, $(d)=2.0$, $(e)=2.2$, and $(f)=10$. The system size is $N=40$. }
\label{fig:APP1}
\end{figure*}
%

%
%
\begin{figure}
\includegraphics[width=0.79\linewidth]{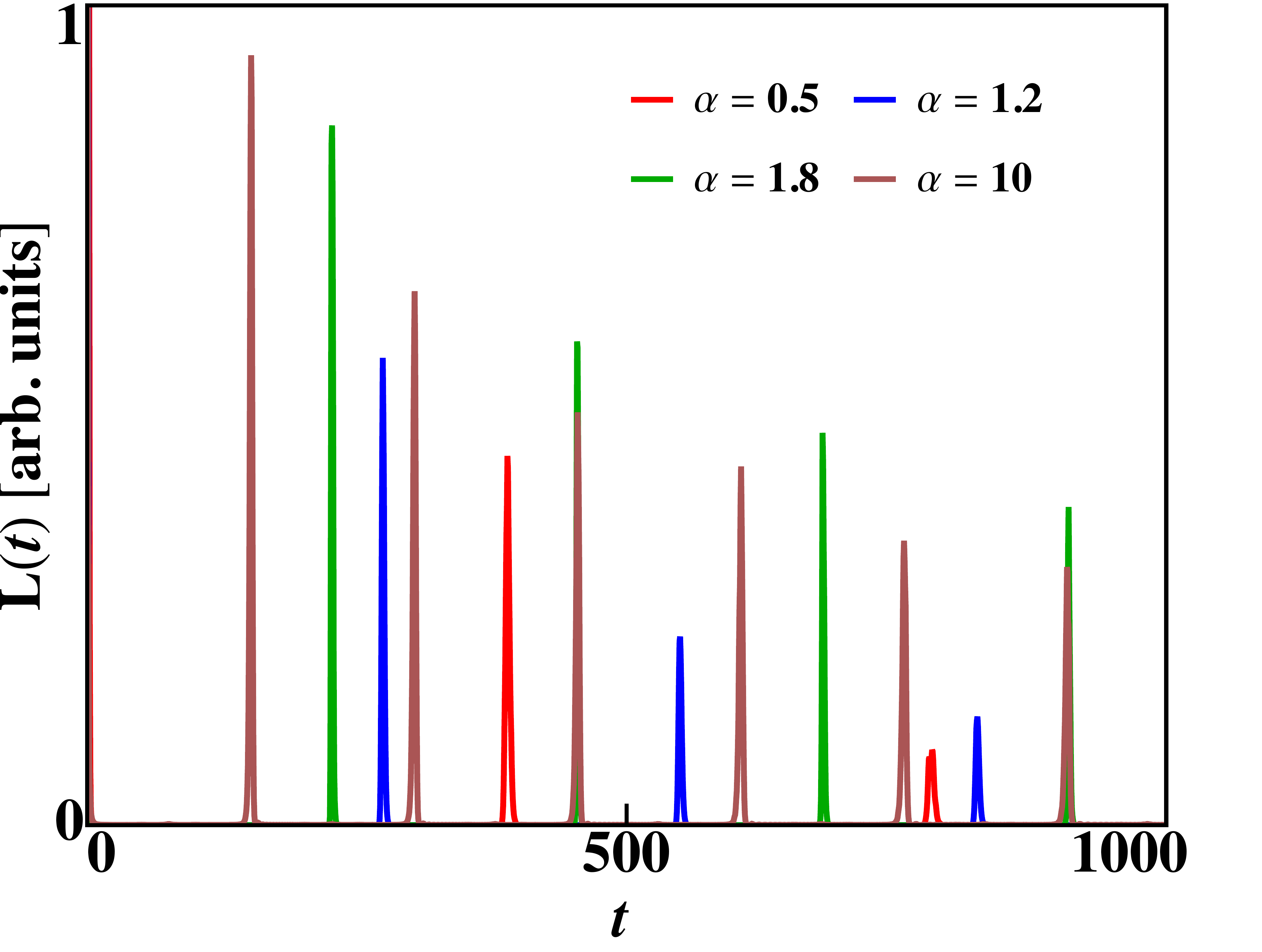}
\vspace{0.cm}
\caption{(Color online) Revivals of Loschmidt echo as in Fig.~\ref{fig:6} but for system size $N=500$.}
\label{fig:APP2}
\end{figure}

%
\begin{figure}
\vspace{-0.5cm}
\includegraphics[width=\linewidth]{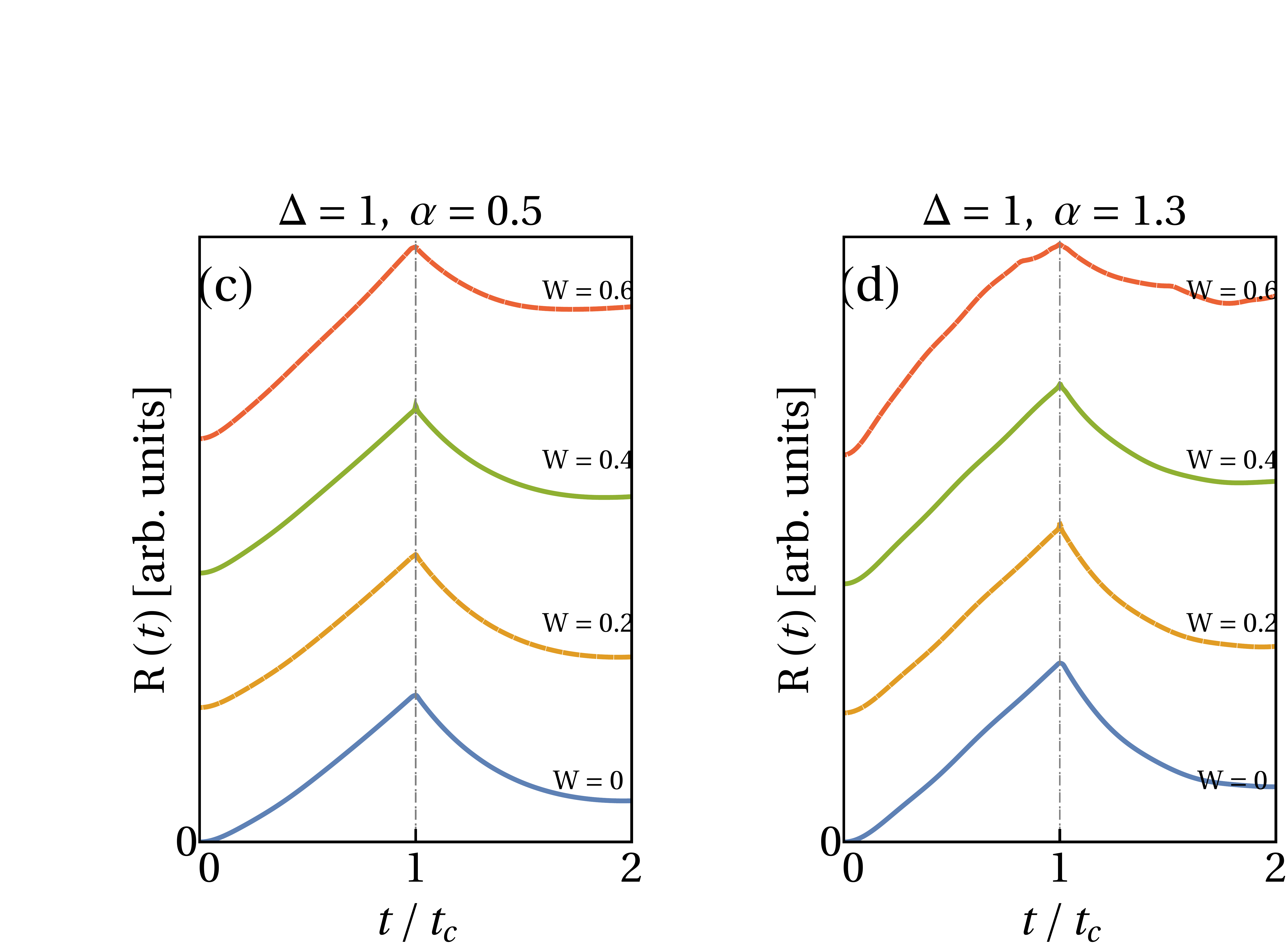}
\\\vspace{-1.15cm}
\includegraphics[width=1\linewidth]{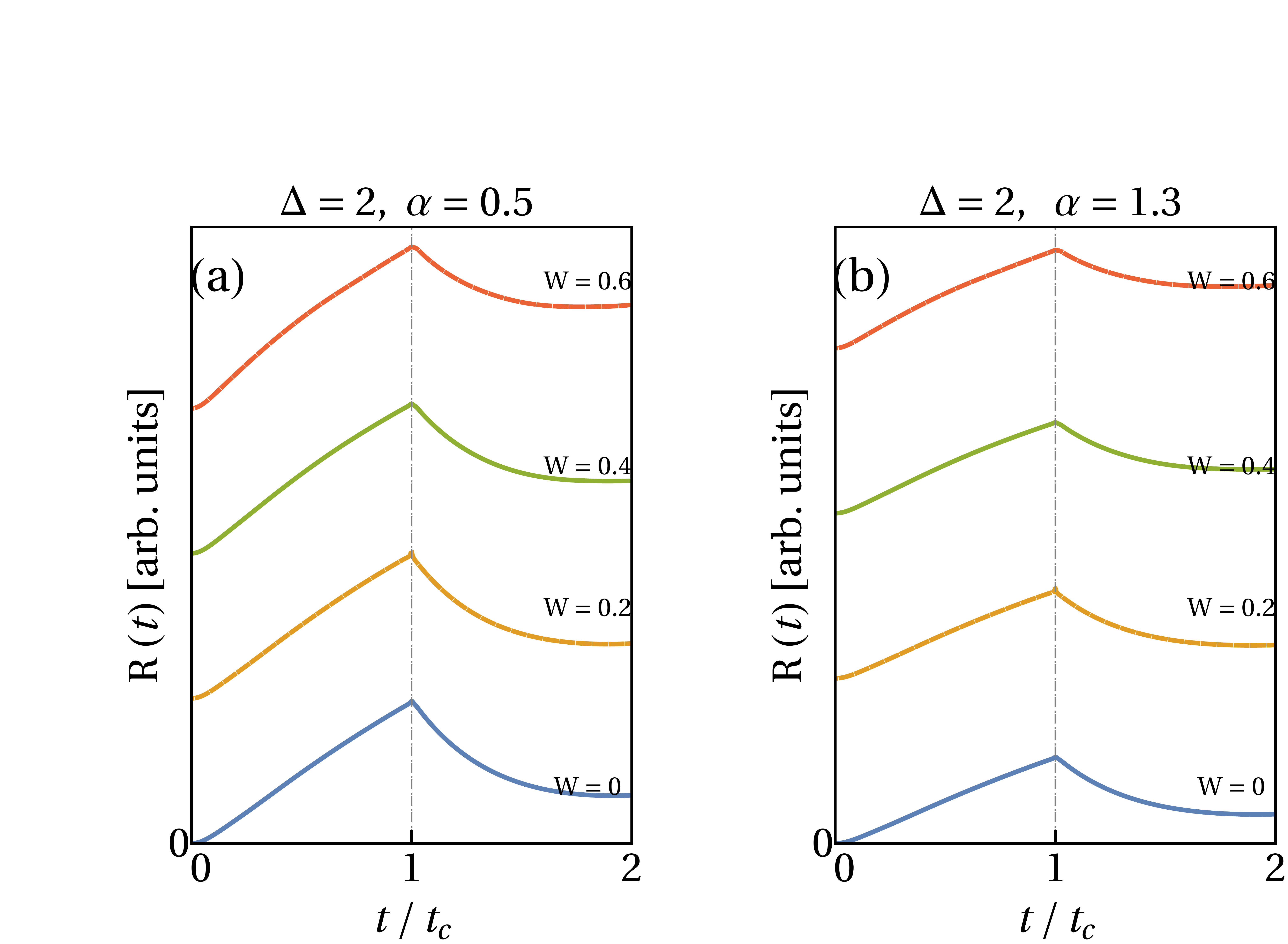}
\caption{(Color online) The rate function of return probability as in Fig.~\ref{fig:10}, but for system size $N=500$. The quenching parameters are same as in Fig.~\ref{fig:10}.}
\label{fig:APP3}
\end{figure}

\section*{Acknowledgments}
U.M. and A.A.  are grateful to Peter Fulde, and Jaeyoon Cho for fruitful discussions. The authors' thanks Henrik Johansson,  Tilen Cadez and Abolfazl Bayat for useful comments.
This work was supported through National Research Foundation of Korea (NRF) funded by the Ministry of Science of Korea (Grants  No. 2017R1D1A1B03033465, and No. 2019R1H1A2039733),
and by the National Foundation of Korea (NRF) funded by the Ministry of Science, ICT and Future Planning (Grant No. 2016K1A4A4A01922028).


%
\appendix

\section{Behvaior of energy spectrum with pairing exponent, $\alpha$}
\label{AppA}
 In the main text, we have shown the spectrum of the long-range Kitaev chain for three different pairing exponents, $\alpha=0.5,1.2$ and $1.8$, using the exat diagonalization of finite size system of linear length $N=500$.  The spectrum shows that the gap in the energy band depends on the onsite energy $\mu$, pairing exponent $\alpha$, and system size $N$. The gap at $\mu=1$ is closed at $\mu=1$ but it is gapped at $\mu=-1$. This feature is also true for small system size. In Fig.~\ref{fig:APP1}, we plot the spectrum for $N=40$ for $\alpha=0.5,1.2,1.8,2,2.2$, and $10$. Here also the gap at $\mu=-1 $ is quite visible while vanishingly close at $\mu=1$. The gap closing point is important to observe the revivals of Loschmidt echo, as discussed in the main text.

%
 \section{Revivals of Loschmidt echo for larger system size}
 \label{AppB}

It is important to understand the robustness of Loschmidt echo for higher system size in order to make  a definite conclusion. For this, we consider $N=500$ and investigate the dynamics of $L(t)$ for quenching from initial ground state of the system. The results of this perusal is reported in Fig.~\ref{fig:APP2} for different values of $\alpha$. The other parameters are same as in Fig. \ref{fig:6}. We see that even for higher system size, there are equally spaced revivals in the Loschmidt echo with decreasing amplitudes of successive revivals.
%

\section{Dynamical phase transitions for larger system size}
\label{AppC}

In this part, we report the dynamical phase transition for system size $N=500$. The plots of the rate of return probability, $R(t)$ for two different values of $\Delta$ and $\alpha$ are shown in Fig.~\ref{fig:APP3}. Each curve in the figures are for different values of disorder strength $W$. From the curves, it is clear that the singularity in $R(t)$ persists for low disorder strength while it get smothen for large disorder strengths.  This observation agrees with the result reported in Fig.~\ref{fig:10}.

%

\bibliographystyle{prsty}
\bibliography{Ref}
\end{document}